# A High-Scale Assessment of Social Media and Mainstream Media in Scientific Communication

Yang Yang[a], Tanya Y. Tian[b,c], Brian Uzzi[d], and Benjamin F. Jones[d,e,*]

Author Affiliations: [a] Mendoza College of Business, University of Notre Dame; [b] New York University Shanghai; [c] New York University; [d] Kellogg School of Management, Northwestern University; [e] National Bureau of Economic Research.

* Corresponding author. E-mail: bjones@kellogg.northwestern.edu

October 2025

**Abstract**
Communication of scientific knowledge beyond the walls of science is key to science's societal impact[1-3]. Media channels play sizable roles in disseminating new scientific ideas about human health, economic welfare, and government policy as well as responses to emergent challenges such as climate change[4-10]. Indeed, effectively communicating science to the public helps inform society's decisions on scientific and technological policies, the value of science, and investment in research[2,11-14]. At the same time, the rise of social media has greatly changed communication systems, which may substantially affect the public's interface with science[11,15]. Examining 20.9 million scientific publications, we compare research coverage in social media and mainstream media in a broad corpus of scientific work. We find substantial shifts in the scale, impact, and heterogeneity of scientific coverage. First, social media significantly alters what science is, and is not, covered. Whereas mainstream media accentuates eminence in the coverage of science and focuses on specific fields, social media more evenly sample research according to field, institutional rank, journal, and demography, increasing the scale of scientific ideas covered relative to mainstream outlets more than eightfold. Second, despite concerns about the quality of science represented in social media, we find that social media typically covers scientific works that are impactful and novel within science. Third, scientists on social media, as experts in their domains, tend to surface high-impact research in their own fields while sampling widely across research institutions. Contrary to prevalent observations about social media, these findings reveal that social media expands and diversifies science reporting by highlighting high-impact research and bringing a broader array of scholars, institutions and scientific concepts into public view.

**Introduction**

The rise of social media, and the decline of traditional media sources, raises new questions about how the public interfaces with scientific ideas [16-21]. Social media coverage of science is based on new technologies that disintermediate traditional media organizations and rapidly disseminate news worldwide [17,22,23]. Yet, our understanding of the scale, quality, and heterogeneity of science presented in social media, and how this compares to science coverage in mainstream media, remains incomplete [7]. On the one hand, concerns exist in many domains about the quality of information on social media [16-19,24-27]. The journalistic norms and regulatory mechanisms that help traditional media outlets balance reporting hype and accuracy appear uncommon when social media disseminates news [5,7,28,29], an issue that may disrupt the adoption of innovation, policymaking, and public trust in science [17,22,30]. On the other hand, social media may rely less on individual and institutional eminence than mainstream media, thereby representing a broader cross section of scientific ideas and promoting potentially important science that stands outside the most prominent institutional settings [31,32]. Social media also enables direct and immediate communication between scientists and the public [23]. Scientists on social media can offer first-hand accounts of recent work, potentially lessening intermediation problems caused by non-expert reporting of complex scientific ideas.

This paper combines large-scale datasets on scientific articles and media coverage. By comparing media coverage of science with the baseline corpus of scientific works, we can examine both what science is covered and what is not covered, allowing us to quantify how reporting rates in mainstream media and social media differ in their focus across scientific papers, fields, institutions, and researchers. Our publications dataset covers 20.9 million journal articles published from 2012 to 2019 as recorded in the OpenAlex database. We integrate this paper-level bibliographic information with data on mainstream media and social media references via the Altmetric database (see **SI 1.2.1**). Altmetric tracks mentions of scientific publications across multiple sources, including mainstream media and social media platforms such as X (formerly Twitter). At the intersection of these databases, we focus on over 699,000 scientific papers covered by more than 4,300 mainstream media outlets (MM) and 5.8 million papers tweeted by 4.6 million Twitter users (SM) associated with our sample of 20.9 million journal articles. The MM outlets include all major newspapers and cable news outlets in the United States (see **SI S1.1.5** and **SI S1.1.8**). For social media, we focus on X/Twitter given its dominant role among social media sources in presenting new scientific papers to the public (see **SI S1.1.6**). To identify the source of social media posts, we use existing methods to match X/Twitter users and scientific researchers [33] as detailed in **SI S1.2.6**. Finally, we replicate our main findings using scientific articles covered on Facebook (see **SI S7**), the number two social media source for mentioning scientific articles (**Table S2**).

*Scale*

We first consider the scale of scientific publications covered by different media (Figure 1). Mainstream media is traditionally viewed as a primary platform for scientific communication [34]. Nevertheless, relative to the 20.9 million scientific publications from 2012 to 2019, we find that reporting through MM outlets is extremely limited, covering only 3.3% of all the publications available. Strikingly, SM increases coverage to 27.5% of all publications, multiplying the scale of unique papers covered by mainstream outlets more than 8 times. Consistent with prior work [35,36], we observe overlap in the specific publications covered by mainstream outlets and social media. As shown in Figure 1a, conditional on being reported by mainstream media, a paper has an 81.8% chance of being covered in social media. By contrast, only 9.8% of scientific papers covered in social media are also covered by mainstream media sources.

Figure 1a further shows that the annual number of scientific publications is growing rapidly, yet the volume of papers covered in mainstream media is stagnant. Mainstream media thus covers a smaller and smaller slice of science. By contrast, the number and share of publications covered in social media is rising. By 2019, 33.2% of science publications are covered on Twitter/X, compared to only 16.7% in 2012 (and zero coverage prior to Twitter's founding in 2006). Assuming that a mainstream media article's depth of coverage and rate of reading are roughly stable over time, these findings indicate that MM's share of science communication is shrinking while social media's share is expanding and progressively dominating MM in reporting rates (see **SI** for details on audience volumes). Figure 1b illustrates differences in the scale of coverage between mainstream media (MM) and social media (SM) based on a semantic analysis of 17 million paper abstracts available for the 20.9 million articles in our corpus. Using a two-dimensional projection of semantic space, the gray points represent the estimated range of topics in our data, blue and pink points represent topics covered by MM and SM respectively. We observe that papers discussed on SM are more uniformly distributed across the semantic space than MM, which is concentrated in specific regions (see **SI S8.1** for methodological details and Figure 3a, **SI S4.1**, for robustness checks). Frame and sentiment analyses on MM articles and SM tweets reporting on science confirm prior studies on MM and SM content differences[5,7,28,29]. MM articles tend to frame research cautiously and over a broad range of positive to negative sentiment, while SM tweets disseminate awareness about research and mostly involve neutral sentiment (see **SI S8.2** and **S8.3** for methodological details).

*Impact and Novelty*

Given SM's growing prevalence in covering science, our second set of analyses investigate the citation impact and innovativeness of the publications covered in social media and mainstream media outlets [37]. We focus on two key measures. First, paper impact is measured through citations received, normalized as the paper's citation percentile among all papers published in a given field and year (see **SI 1.2.4**). Alternative impact measures are presented in the **SI**. Second, building on prior literature [38-42], we estimate a paper's novelty in science based on the degree to which a paper combines prior knowledge in unfamiliar ways. Papers that combine prior work in statistically uncommon ways are considered more novel (see **SI S1.2.3**). The novelty measure is presented in percentiles, with higher values indicating greater levels of novelty (see **SI 1.2.3**). Notably, impact and novelty pertain to distinctive perspectives on paper performance. Novelty concerns the paper's innovativeness in its use of past knowledge, while impact considers its influence on future research [37].

Figure 2a examines paper impact. Mainstream media heavily emphasizes high impact works. The most highly cited works within science – those in the upper 10 percent of citations - are over-represented by a factor of 3.8 in mainstream media compared to their baseline rate. More generally, mainstream media reporting emphasizes works above the 75th percentile of the citation distribution. Social media also emphasizes high-impact works compared to the baseline but more evenly covers highly-cited papers than MM, emphasizing works above the 60$^{th}$ percentile of the citation distribution. Social media thus significantly expands the volume of science covered (Figure 1) while stressing coverage of high impact work (Figure 2). Among scientific papers at the 90$^{th}$ percentile of citations, SM multiplies coverage over mainstream media by a factor of 3.5. Notably, papers with no citations are heavily under-represented in both social media and mainstream media (Figure 2a, inset).

The orientation towards high-impact works appears when media coverage is either retrospective, engaging publications from earlier years, or contemporaneous within the publication year, suggesting that both mainstream media and social media surface new scientific publications of disproportionately influential future impact (see **SI S2.1.1 & Table S5**). The findings are also robust when weighting by the circulation for mainstream outlets and the number of followers of posts on social media (see **SI S3.1** and **Figure S1b**) or when self-promoting tweets are excluded (see **SI S2.1.3**, **SI S3.1**, **Figure S2c**). Finally, robustness checks indicate that the findings hold in multivariate regressions with wide-ranging controls (**SI S2.1**). All told, the orientation on high-impact works appears among new and old scientific articles and when accounting for estimated audience size and self-promotion.

We next consider novelty. Figure 2b plots the novelty percentile of papers covered by MM and SM. Both media channels emphasize novel papers, compared to baseline rates in science. Papers exceeding the 75th percentile of novelty are over-represented in both MM and SM. However, as with citation impact, social media appears as an intermediate case, favoring novel papers compared to baseline rates but not to the same degree as mainstream media. Regression evidence (see **SI 2.2**, **Table S8**) further supports these findings net of numerous controls, as well as when we consider contemporaneous media coverage (see **SI 2.2.1**, **SI S3.2**, **Figure S4a**), retrospective media coverage (see **SI 2.2.2**, **S3.2**, **Figure S4b**), variation in audience size of the media platforms (see **SI Figure S3**), and when self-promoted tweets are excluded from the analysis (see **SI 2.2.3**, **SI S3.2**, **Figure S4c**).

*Heterogeneity of Coverage*

We further consider heterogeneity in scientific coverage, comparing SM and MM with respect to disciplines, institutions, journals, and demography. Figure 3a examines research fields. We compare SM (red) and MM (blue) coverage, with reporting rates for each field normalized by the number of publications in that field. Fields with the highest reporting rates in MM – psychology, biology, geography, and medicine – also are emphasized in SM. At the same time, reporting rates across disciplines decline less steeply for SM than for MM, indicating SM's relatively broader coverage of research fields (see **SI S4.1** for robustness tests). This broader reach in field engagement by SM is striking for fields such as math, art, and philosophy that appear significantly less in MM.

Figure 3b examines research institutions. We present reporting rates for publications from different universities, organizing universities into bins according to the U.S. News Best Global Universities ranking (see **SI S1.1.2**). Mainstream media (blue line) coverage is seen to strongly favor elite institutions: The higher the rank of the university, the higher the probability of a scientific paper's coverage. Social media (red line) also favors higher-ranked institutions in reporting rates. However, the relatively flatter decay for SM across changes in institutional rank indicates that social media reports comparatively more evenly than MM (see **SI S4.2** for robustness tests).

Figure 3c further presents cumulative distribution functions across universities for publication counts, citation counts, and coverage by MM and SM. Highly ranked universities, compared to both their publications (light grey line) and citations (dark grey line), are disproportionately likely to see their research reported in MM (blue line). Indeed, while the top 50 universities produce 21.3% of all publications and receive 31.3% of all citations, their work receives 42.7% of mainstream media coverage. For SM, the top 50 institutions receive 30.5% of social media coverage. Social media's orientation on highly ranked institutions further increases when weighted by the intensity of attention (see **SI S4.2.1**) but continues to remain less skewed than coverage by MM. Regression analysis confirms that both MM and SM are biased towards highly ranked institutions (see **SI S4.2.1** and **Figure S13-S16**) and that institutional rank strongly predicts coverage after controlling for a paper's bibliographic variables, including citations and novelty.

Figure 3d examines scientific journals. Journals are ranked according to their Scimago Journal Rank (see **SI S1.1.3**) and binned in groups. The findings are broadly similar to the analysis across research institutions. Both media channels strongly emphasize highly-ranked journals. However, social media provides higher scale coverage of research across journals and presents somewhat more egalitarian coverage of journals of varied ranks. These findings are further confirmed when looking at both contemporaneous and retrospective media mentions of scientific papers and in regression analyses with numerous controls (see **SI S4.3** and **Figure S18-21**).

Finally, we consider team demographics. Given the rise of team science and heterogenous team configurations [37,43], we study reporting rates classifying teams as all women teams (aw), all men teams (am), or mixed teams, where the team's first and last author is reported as both men (mm), man-woman (mw), woman-man (wm), or both women (ww). The y-axis indicates the predicted margins for media reporting from fixed effect regressions (see **SI S5**), accounting for a given individual's tendency to be reported upon. Figure 3e indicates that mainstream media de-emphasizes the research of all women teams ($p<.001$) or teams that have women in both first and last author roles ($p<.001$). By contrast, social media shows substantially smaller report rate declines for all women teams and more generally reports more evenly across other team configurations.

**Scientists as Communicators**

Our final analyses consider the role of scientists as communicators and the role they may play in the scale, impact, and range of science that appears through social media channels. Scientific reporting in MM has traditionally worked through intermediaries – journalists and their media institutions. Social media can allow practicing scientists themselves to bring science into the public square, and many academic researchers have developed large SM followings.

We matched scientists in OpenAlex with Twitter users who mention scientific papers (see **SI S1.2.6** for methodological details). This analysis locates a set of 477 thousand scientists who tweet (see **SI S1.2.6**). These scientists collectively posted 16.3 million tweets/retweets mentioning specific scientific articles. Notably, most retweets of scientists' tweets come from members of the public (see **SI S.1.2.7**), indicating that scientists who tweet are communicating scientific works into broader public discourse.

Figure 4 analyzes tweeting scientists and the papers they communicate. First, Figure 4a indicates that, regardless of their number of followers, scientists who tweet are more highly cited on average than other scientists in their same field and cohort. Average citation impact rises from 5 times the field-cohort mean at low follower counts to 15 times the field-cohort mean at the highest follower counts. Turning to the papers scientists tweet, we again see strong links with impact (Figure 4b). Scientists tweet about works that tend to be highly cited compared to other papers in that same field and year, and this tendency toward high impact work increases further among scientists with the largest numbers of followers.

Examining the fields covered by tweeting scientists, Figure 4c shows that scientists tend to tweet about science papers in their area of expertise but their coverage is less confined to research that originates at their own institution. Figure 4d indicates that on average, over 75 percent of scientists' tweets on scientific publications cover research that originates from universities other than their own university. All together, we see that scientists tend to communicate high-impact science, drawing on their area of expertise, and with panoramic attention to a wide set of institutions.

**Discussion**

The reporting of science into the public square, long rooted in traditional journalistic practices and media outlets, has been cast into flux by the rapid changes in the media landscape. Here we examined how a new media channel, social media, compares with traditional media in scientific coverage. By comparing media coverage against the baseline corpus of scientific works, we can examine both what is reported and what is not reported, revealing how reporting rates of science, scientists, and science institutions have changed as the media landscape has evolved.

We find substantial shifts in the scale, impact, and heterogeneity of scientific coverage. Mainstream media surfaces a small and declining share of scientific articles, focusing on high impact and novel works. Further, mainstream media heavily emphasizes science from elite institutions, high-impact journals, specific fields, and research teams led by men. By contrast, social media has significantly expanded the scale of scientific coverage compared to mainstream media. Yet, despite a huge increase in scale, social media maintains an emphasis on high impact and novel scientific research. Social media further presents a broader distribution of scientific works across research fields, institutions, journals, and team configurations.

In shifting away from the journalistic norms, fact-checking, and regulatory arrangements meant to support accuracy in mainstream media, the advent of social media has raised widespread concerns about the quality of information presented [16-19,24-27]. Indeed, social media appears rife with low quality and outright false information in many spheres [10,11,17,18]. However, our findings indicate that social media in the scientific domain tends to focus on high-quality research, particularly in terms of novelty and impact within science itself. In interpreting this emphasis on high-quality research, a key feature may be the widespread presence in social media of scientists themselves, acting as experts who source impactful and novel ideas at high scale. Additionally, while voices on social media need not follow norms of professional journalism, scientists have their own disciplinary norms, Mertonian norms of science [44] emphasize truth-seeking and organized skepticism, and these may serve in part as substitutes for traditional journalistic practices. Scientists' expertise, as well as their reputational concerns, may then be significant forces for sourcing high-quality content in their presentations on social media. Science in social media may thus exhibit distinct information quality from other social media domains, such as general news, given the presence of scientists and their parallel norms.

While social media in science emphasizes high impact works, it also increases the heterogeneity of disciplines, institutions, and scientists covered. Social media may then surface key ideas and researchers that are otherwise overlooked, a salient feature given that the history of science offers many cases where ideas were long ignored and emerged outside the elite [45,46]. Further, in the science context, the broadening range of coverage across ideas, fields, institutions, and individuals compared to mainstream media may act to lessen the Matthew Effect, a well-known "rich get richer" phenomenon where elite scientists and institutions are known to gain disproportionate attention and resources [47-50]. Social media may then widen avenues to public appreciation and opportunities across the science landscape. That said, while social media proves more egalitarian than mainstream media on many dimensions of scientific coverage, we still see biases in social media, including toward elite institutions. Social media's role in broadening coverage thus appears incomplete.

Our analysis is limited in several ways. As a first large-scale analysis comparing different media platforms against the broad corpus of scientific work, we focused our analysis on the differences in coverage of topics, institutions, journals, and team demographics. Future studies could analyze additional dimensions, including individual level characteristics such as career stage and nationality. Further, this study mainly focuses on rates of reporting, not the content or depth of coverage, which is an important avenue for future work that can potentially harness new AI methods to further inform and enhance science's communication ecosystem. Finally, while our analysis uses high scale data, it naturally focuses on specific communication channels. Future research may extend attention to other communication channels to further examine additional means by which science is presented to the public.

Communication of science is key to an informed public [3,11]. It is also central to appreciation of science as a public good and support for investment in science [12,14,51]. Overall, our findings suggest that social media is substantially shifting scientific communications, expanding the reach of science into the public square, surfacing more varied works and sources, while emphasizing especially novel and impactful work within science itself.

## Method
Data and methods used in our work are described below.

### Data Sample
Our primary dataset is derived from OpenAlex (https://openalex.org/), a comprehensive open-access bibliographic database of scholarly publications. For each journal article, OpenAlex provides metadata including bibliographic details (e.g., journal, volume, issue, page numbers, and publication date), authorship information (e.g., author names, affiliations, and disambiguated identifiers), journal classification, field, and citation linkages to other indexed articles. We focus on 20.9 million journal articles published between 2012 and 2019 in venues recognized by the Scimago Journal Rank (SJR) (see **Section S1.1.3** for details). To compare research covered in social media and mainstream media, we integrate article-level metadata from OpenAlex with media attention data from the Altmetric database (accessed on June 3, 2021). The Altmetric platform tracks mentions of scientific publications across a wide range of online sources, including both mainstream media outlets (e.g., CNN, MSNBC) and social media platforms (e.g., X/Twitter, Facebook), thereby enabling a comprehensive analysis of science communication.

A key component of our analysis examines heterogeneity in scientific media coverage across multiple dimensions, including disciplinary fields, institutional affiliations, journal prestige, and author demographics. To support this investigation, we augment our main dataset with additional attributes: institutional rankings, journal rankings, and author gender information.

**U.S. News University Ranking.** Our institutional ranking data comes from the U.S. News Best Global Universities Rankings (https://www.usnews.com/), collected on September 29, 2021. Their rankings include approximately 1,500 universities across 80 different countries. For institutions that are not listed in U.S. News website, we classify them into a category called "unranked" in regression analyses.

**Scimago Journal Rank.** To focus on high-impact outlets, we matched OpenAlex with journals listed in the Scimago Journal Ranking (SJR; https://www.scimagojr.com). The SJR evaluates journals based on citation data, weighting citations from highly ranked journals more heavily than those from lower-ranked ones. We matched the SJR list with OpenAlex using journal names, International Standard Serial Numbers (ISSNs), and the ISSN-L table (https://www.issn.org/). This procedure matched 30,894 journals (90.0%) in the SJR to their OpenAlex counterparts. For those not automatically matched, we manually identified an additional 275 journals. In total, 31,169 of 34,337 SJR-listed journals (90.8%) were linked to OpenAlex.

**Gender Inference of Authors.** Following our previous work[37], we used the NamSor API (https://namsor.app/) to infer the gender of authors in our sample. NamSor API v2.0.26 was used to infer the gender of authors in our sample. The algorithm was run on the first and last names of authors in our sample. This procedure employs a binary gender system, consistent with most existing gender studies in science. Among scientists with both first and last names available, 58.7% were classified as men and 41.3% as women (see **Section S1.2.2** for details). Because the method demonstrated face validity in our earlier study of team gender composition[37], we did not repeat robustness checks here.

**Mainstream Media Circulation and Twitter Follower.** We tested the robustness of our findings by weighting mainstream outlets by estimated circulation and social media posts by user followership. Circulation was proxied using backlink data from Majestic SEO (https://majestic.com/; accessed November 3, 2024), which records the number of external websites linking to a given domain. All 4,907 media outlets in the Altmetric database had corresponding backlink data in Majestic. For X/Twitter, we weighted mentions by follower counts of the posting accounts. Altmetric provides identifiers for both the tweeter and the tweet when scientific outputs are mentioned, enabling our collection of detailed data on ~10.2 million users and 126 million tweets via the Twitter API (https://developer.twitter.com/en/docs/twitter-api/). For each user, we extracted identifiers, handles, display names, profile and location descriptions, follower and following counts, tweet totals, account creation dates, and verification status. For each tweet, we collected identifiers, text, engagement metrics (retweets, likes, replies, quotes), timestamps, and references to other tweets.

**Measures**

The important measures used in our analysis are presented as below.

**Impact measures.** The OpenAlex database records annual citation counts for each paper. We define "*hit papers*" as those in the top 5% of citations within their field and publication year, coded as a binary indicator. As an alternative measure of impact, we normalizing a paper $i$'s final citations by the field-year average, which is denoted as $\hat{c}_i$ (see **Section S1.2.4**). For robustness, we estimate regressions using both the binary *hit paper* variable and the continuous normalized measure as dependent variables. Given that $\hat{c}_i$ follows a heavy-tail distribution, we consider a log transform of $\hat{c}_i$,

$$impact = \log(\hat{c}_i + 1) \tag{1}$$

For the analyses in **Figure 2**, we convert the continuous impact measure into percentiles, with higher values indicating greater paper impact.

**Novelty measures.** Following prior research[42], we measure a paper's novelty as the extent to which a paper contains statistically atypical combinations of references. The metric is based on a $z$-score that compares the observed frequency of journal pairings in a paper's reference list to the expected distribution generated from randomized citation networks. Journal pairings with $z$-scores below zero are considered novel, reflecting uncommon combinations of prior work, whereas those with scores above zero are considered conventional. Therefore, each paper has a distribution of $z$-scores for all journal parings in its reference list. The novelty measure of a paper is defined as the tenth percentile value of this distribution, where lower values indicate higher novelty. Details of the method can be found in pages 3 – 5 in the supplementary information of work[42].

As with impact, we define both binary and continuous measures of novelty. A paper is coded as *novel paper* (binary) if its $z$-score is below zero. For the continuous measure, we apply an inverse log transformation of the $z$-score so that higher values correspond to greater novelty in our regression analysis in **Section S2**. In **Figure 2**, we further convert this measure into percentiles to aid interpretation, with larger percentiles indicating higher novelty.

**Team gender composition.** To analyze media coverage by team demographics, we classified teams into six categories based on identified gender of authors: all women team (aw), all men team (am), mixed teams with both first and last authors women (ww), mixed teams with a man first author and woman last author (mw), mixed teams with a woman first author and man last author (wm), and mixed teams with both first and last authors men (mm). We denote team composition variable as $G_i \in \{aw, am, ww, mw, wm, mm\}$.

**Regression Analysis**

The results in **Figure 2** and **Figure 3** are validated using fixed-effects ordinary least squares regressions. Models control for potential confounds including authorship, team size, leadership, institutional prestige, publication year, journal quality, prior citation impact of authors, and average team career age (see **Section S2.1**). To test robustness in analyses of team gender composition presented in **Figure 3f**, we additionally include individual fixed effects in regression analysis (see **Section S5**).

As shown in **Figure 3**, social media (X/Twitter) provides more egalitarian coverage of scientific papers than mainstream media across disciplines, institutions, journals, and team gender compositions. To validate this pattern, we estimated fixed-effects OLS regressions predicting the likelihood of mention in Twitter and mainstream media. From these models, we calculated predicted margins for field, institutional rank, journal rank, and team gender composition fixed effects. Results reported in **Sections S4** and **S5** confirm Twitter's more egalitarian coverage even after controlling for numerous confounding factors.

**Matching Scientists on Twitter**

We analyze Twitter accounts tracked by Altmetric (as of June 3, 2021) that mentioned scientific articles, focusing on cases where researchers promoted their own work. Following Mongeon et al. [33], we assume scientists active on Twitter often disseminate their publications. To identify self-promotion, we extract names from usernames and display names, removing common prefixes and suffixes (e.g., "Dr.," "PhD"). Different from prior work[33], we also account for nicknames—e.g., mapping "Ben" to "Benjamin"—using a nickname dictionary for more accurate matching. Our matching process first links full first and last names between tweeters and authors in OpenAlex. When only an initial is available, we match by using first initial and last name. This matching procedure yields 484,032 tweeters linked to 499,440 OpenAlex authors. These statistics are consistent with findings reported in the work of Mongeon et al. [33]. For ambiguous cases, we refine matches using institutional affiliation, geographic location, research field, and co-authorship or retweeting patterns. This curation produces 476,890 unique tweeter–author pairs. Restricting to authors whose first publications appeared between 1980 and 2020, our main analysis focuses on 386,000 unique pairs (**Figure 4**).

**Mapping Scientific Papers in a Semantic Space**

To visualize the scale and breadth of scientific ideas represented in mainstream media and social media, we examined the semantic structure of the full corpus of scientific abstracts (see **SI S8**). Using a word2vec model trained on 33 million abstracts published between 2000 and 2022, we embedded 17 million papers from our sample into a 200-dimensional semantic space and projected them into two dimensions using incremental PCA and t-SNE. The resulting map (Figure 1b) highlights the expansive reach of social media relative to mainstream media. Papers mentioned on social media are broadly and evenly distributed across the semantic landscape of science, whereas mainstream media coverage remains concentrated within a few clusters. These text-based patterns reinforce the quantitative evidence presented in Figure 1 and Figure 3: social media substantially broadens both the scale and the topical diversity of scientific ideas reaching the public sphere.

**Data availability.** The publication metadata used in this study derive from the OpenAlex database (https://openalex.org). Media coverage data were obtained from the Altmetric, and X/Twitter data were collected via the Twitter API. Requests for access to underlying third-party data may be addressed to the authors, subject to data-provider approval and applicable data-use agreements.

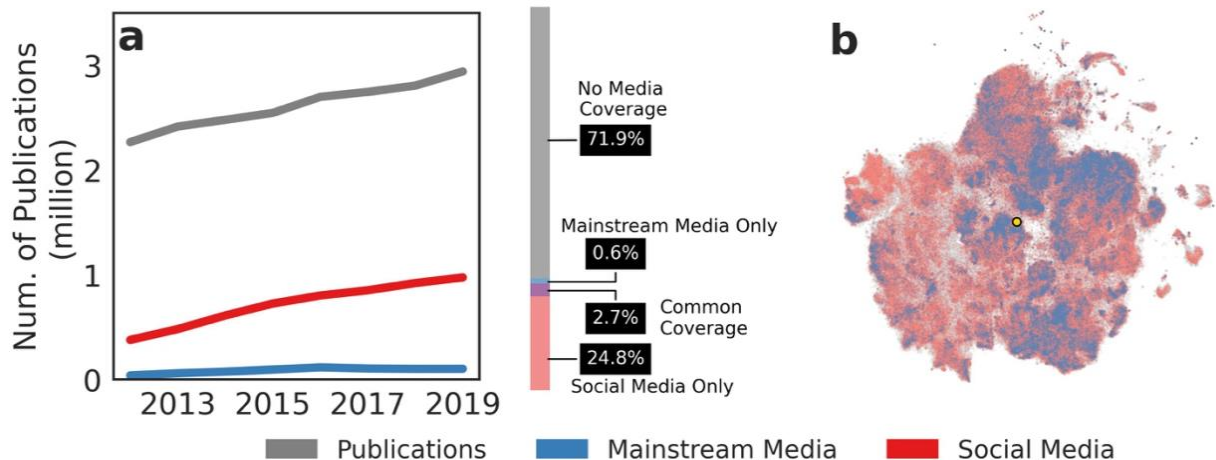

**Figure 1 Scale of Mainstream Media and Social Media Coverage.** (**a**) There is a sharp upward trend of scientific publications (journal articles) from 2012 to 2019 (gray line). Mainstream media coverage (blue line) is limited and report rates have recently declined. The chance of being reported by mainstream media is 3.7% in 2015 and 3.4% in 2019. Social media (X/Twitter) mentioned articles are rising (red line) and gaining as a share of scientific publications. The Venn diagram on the right shows that 28.1% of scientific publications are covered by either mainstream or social media. Of these, 0.6% (light blue) are reported exclusively by mainstream media, 2.7% (light purple) by both, and 24.8% (pink) exclusively by social media. Among papers featured in mainstream media, 81.8% also appear on social media, whereas only 9.8% of papers covered by social media are reported by mainstream media. (**b**) A two-dimensional semantic map of scientific papers in our sample (light gray), those covered by mainstream media (blue), and those covered by social media (pink; X/Twitter) illustrates the scale difference between the two forms of coverage. Papers mentioned on SM (pink points) are widely distributed across the full corpus (gray points) relative to those reported by MM (blue points), indicating that SM covers a more diverse topic space semantically relative to MM, which is more concentrated in specific topic areas. Of the 20.9 million articles analyzed, approximately 17 million include abstracts in OpenAlex. We embedded these papers in a 200-dimensional vector space using a word2vec model trained on ~30 million abstracts published between 2000 and 2022 (see **SI S8.1**), and projected them into two dimensions using incremental PCA for dimensionality reduction, followed by t-SNE for visualization. For clarity, the figure displays a 10% random subsample of all papers, along with those covered by mainstream and social media.

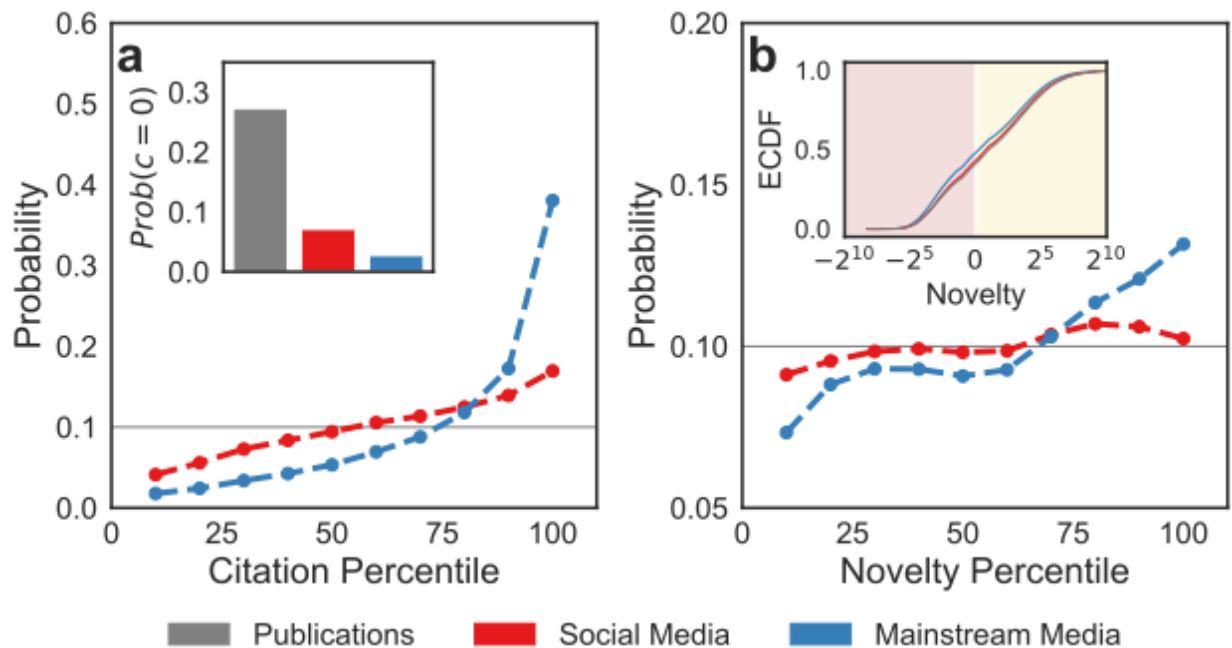

**Figure 2 Paper Impact and Novelty in Media Coverage**. (**a**) Paper impact is converted to percentiles within the paper's field and publication year. Gray/red/blue lines represent relative probabilities of paper impact for all journal articles/social media reported articles/mainstream media reported articles. The inset figure reports the percent of papers with zero citation among all journal articles (gray bar, 27.3%), social media reported articles (red bar, 7.1%), and the mainstream media reported articles (blue bar, 2.7%). (**b**) Paper novelty measures the degree to which a paper combines past knowledge in a new way [42], which is also converted to percentiles. Large percentile values indicate high levels of novelty. Gray/red/blue lines represent the probability density distributions of paper novelty for all journal articles/social media reported articles/mainstream media reported articles. The inset figure represents the novelty $z$ score's [42] cumulative density distribution, where negative z scores indicate novel combinations of past knowledge, while positive z scores indicate conventional knowledge combinations.

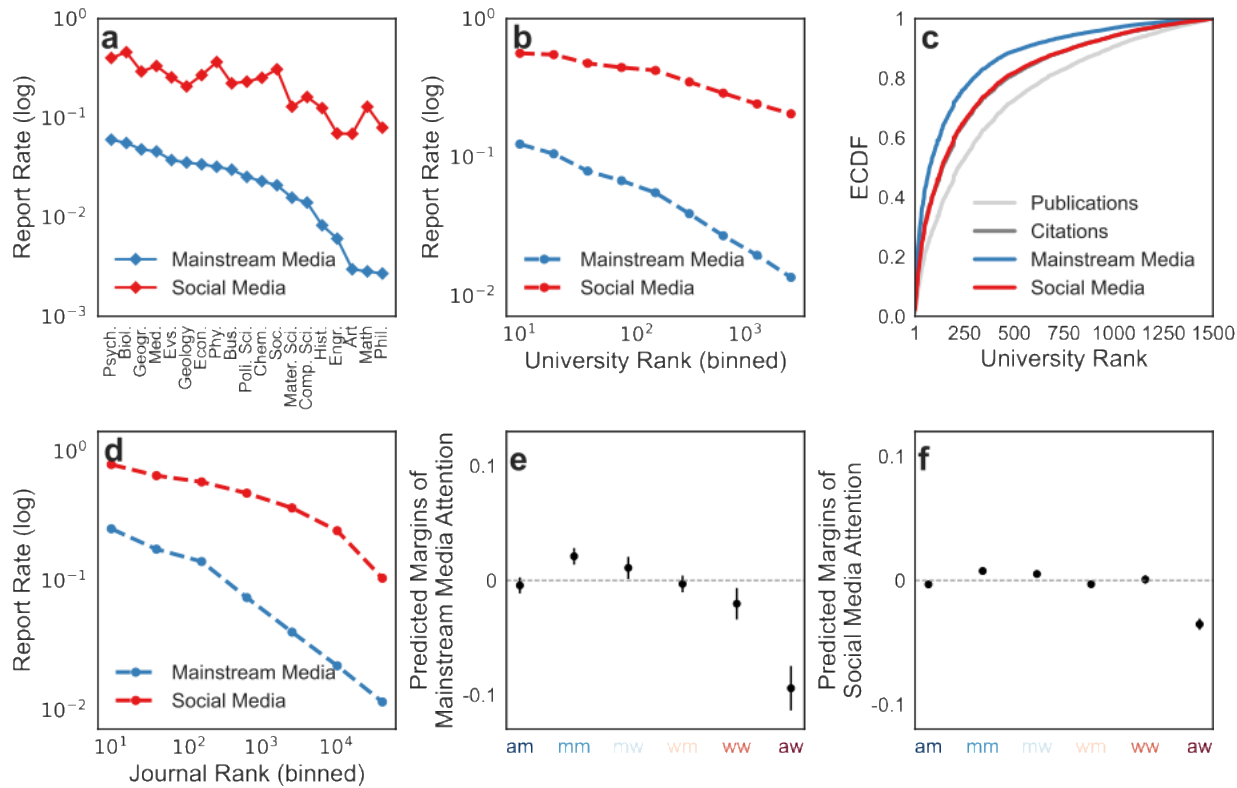

**Figure 3 Diversity in Media Coverage.** (**a**) Comparing across the 19 major research fields, the probability a given paper is reported in social media (red) and mainstream media (blue). (**b**) Report rates by institutional rank and media type. (**c**) Cumulative distributions by institutional rank comparing publication counts (light gray), citation counts (gray), social media reports (red) and mainstream media reports (blue). (**d**) Report rates by journal rank and media type. (**e** and **f**) Media attention based on the research team's demographic composition, comparing teams with all men (am), all women (aw), and mixed teams, where the team's first and last author are both men (mm), man-woman (mw), woman-man (wm), or both women (ww) (see **SI S5** for regression specification and control variables). (**e**) Mainstream media report rates. (**f**) Social media report rates.

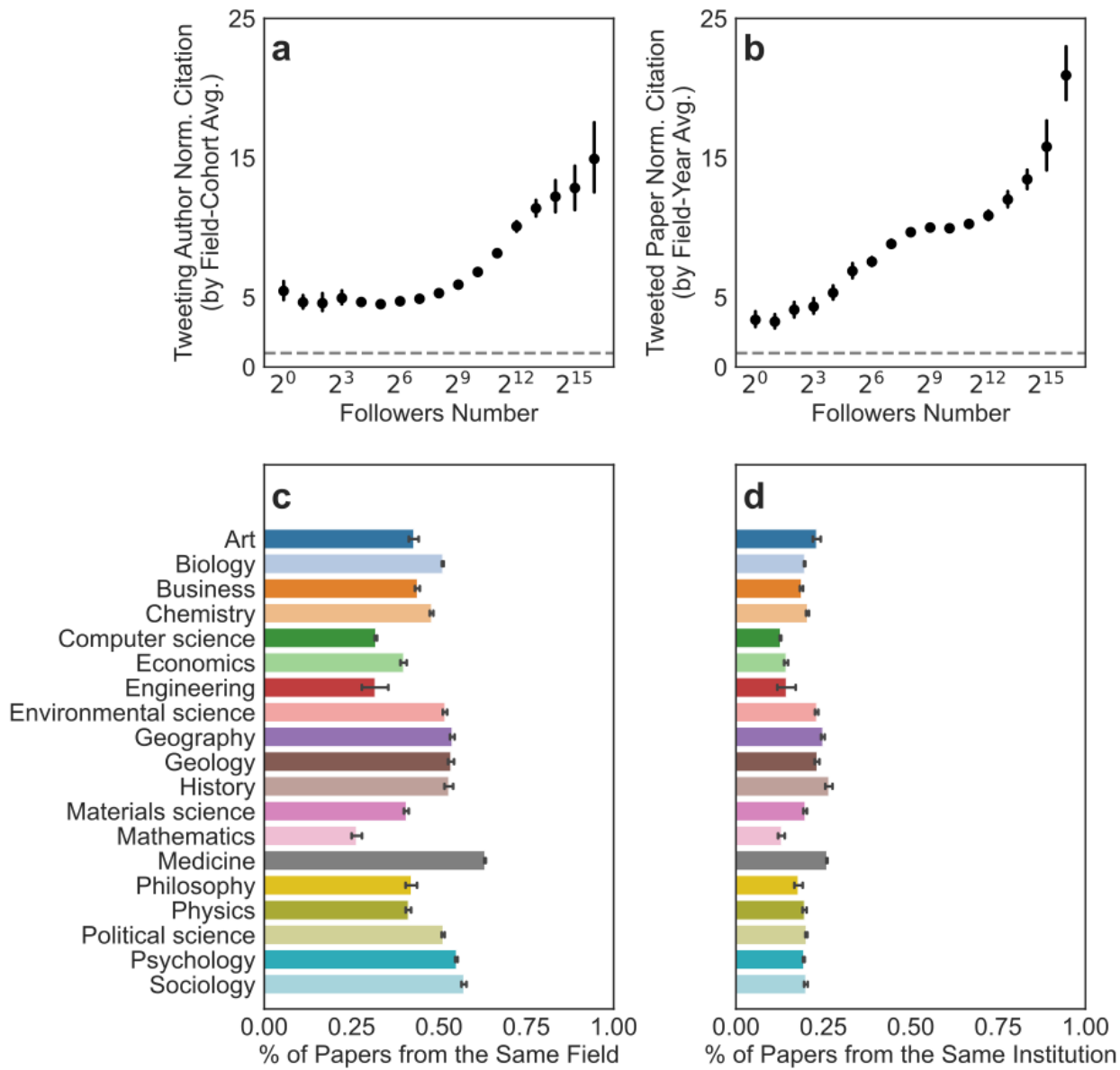

**Figure 4 Scientists on Twitter**. We match scientists who are active on Twitter with their publications in OpenAlex. (**a**) Scientists on Twitter have high mean citations compared to their field-cohort average, and a scientist's follower count is positively correlated with the scientist's citation count. (**b**) Scientists on Twitter report relatively high-impact works, and scientists with higher number of followers tweet papers with higher impact. (**c**) Scientists' tweets often consider papers in the scientist's own primary field. (**d**) Scientists' tweets are much less likely to focus on papers from their own institution.

# Supplementary Information

A High-Scale Assessment of Social Media and Mainstream Media in Scientific Communication

Yang Yang[a], Tanya Y. Tian[b,c], Brian Uzzi[d], and Benjamin F. Jones[d,e,*]

Author Affiliations: [a] Mendoza College of Business, University of Notre Dame; [b] New York University Shanghai; [c] New York University; [d] Kellogg School of Management, Northwestern University; [e] National Bureau of Economic Research.

October 2025

**This PDF file includes:**

    Supplementary text
    Figures S1 to S39
    Tables S1 to S15
    SI References

# 1 Materials and Methods

## 1.1 Data

Our main data sources are described in detail below.

### 1.1.1 OpenAlex Database

We use publication data from OpenAlex (*1*), a comprehensive database of scientific articles. OpenAlex provides, for each article, bibliographic details (including journal, volume, issue, page numbers, and publication dates), authorship records (including names, affiliations, and disambiguated author identifiers), journal information, field information, and citation links to other articles in the database. Our study focused on articles published in journals recognized by the *Scimago Journal Rank* (SJR) (see **Section 1.1.3** for detail), and on publications from the years 2012 through 2019. OpenAlex is publicly available via the link: https://openalex.org/.

### 1.1.2 U.S. News University Ranking Data

Institutional ranking data comes from the U.S. News Best Global Universities Rankings (https://www.usnews.com/education/best-global-universities/rankings), accessed on September 29, 2021. There rankings include approximately 1,500 universities across 80 different countries. The rankings are calculated using 13 weighted indicators to measure a university's global research performance. Highly-weighted factors include global research reputation (12.5%), regional research reputation (12.5%), publications (10%), total citations (12.5%), and the number of publications that are hit papers (12.5%), among other characteristics. For institutions that are not listed in U.S. News University ranking, we classify them into a category called "unranked" in regression analyses. These data are publicly available on the U.S. News website.

### 1.1.3 Scimago Journal Rank

To focus on a large set of relatively impactful journals, we matched OpenAlex with journal identifiers from the Scimago Journal Ranking (SJR). The Scimago Journal Ranking (https://www.scimagojr.com) encompasses 34,337 journals from 1999 to 2020. The SJR ranks academic journals based on citation data, considering both the quantity of citations and their source. Unlike Impact Factor, which treats all citations equally, citations from highly-ranked journals contribute more to the SJR score than those from lower-ranked journals.

The Scimago journal list was matched to OpenAlex journals using journal names, International Standard Serial Numbers (ISSN), and the ISSN-L table (accessed from https://www.issn.org/). On this basis we successfully matched 30,894 (90.0%) of the journals in the SJR with journals documented in OpenAlex. Furthermore, for journals in the SJR that were not automatically matched in the step above, we manually matched 275 additional journals with their counterparts in OpenAlex. In total, 31,169 (90.8%) out of 34,337 journals documented in the Scimago Journal Ranking were matched with the journals recorded in OpenAlex.

### 1.1.4 Altmetric Database

The Altmetric database (accessed June 3, 2021) records scientific outputs and their associated mentions online. These mentions include mainstream media coverage, social media coverage, citations that appear in Wikipedia and public policy documents, and discussions on research blogs, among other sources (**see Table S1**). We incorporate mainstream media mentions, Twitter mentions, and Facebook mentions, all available online. The work of (*2*) has found that the vast majority of news outlets had already established a digital presence by 1996. Scientific outputs tracked by Altmetric are those that: a) are assigned a permanent identifier, such as DOI and PMID; b) are mentioned in at least one "consuming" source (see **Table S1**); and c) are published in the whitelist of journals that Altmetric tracks. The merging of Almetric with broader publications data is discussed below in **S1.2.1**.

### 1.1.5 Mainstream Media Circulation

To estimate the circulation of various mainstream media outlets, we extracted data from Majestic SEO (https://majestic.com/), accessed on November 3, 2024. Majestic records the number of external websites linking to specific domains. For example, *The New York Times* has approximately 32,205 backlinks to nytimes.com, significantly more than *The Chicago Tribune,* which has 3,237 backlinks. According to the work of (*3*), the linking-in sites can be used as a good proxy for the circulation of mainstream media outlets. All 4,907 media outlets from the Altmetric database have corresponding backlink data in Majestic.

**Table S1 Sources in Altmetric Database (accessed as of June 3, 2021).**

| Source | Description |
|---|---|
| Mainstream | 4,000+ English and non-English news outlets |
| Social media | X (formerly Twitter), Facebook, reddit, google+, etc. |
| Blogs | 14,000 academic and non-academic blogs |
| Patents | Patents from Australia, Germany, Switzerland, European Patent Office, United States, France, United Kingdom, and Netherlands |
| Policy documents | Policy, guidance, or guidelines document from government or non-government organizations |
| Wikipedia | Wikipedia website |
| Q&A website | StackOverflow website |
| Online reference managers | Mendeley and Citeulike |
| Post-publication peer-review forums | Pubpeer and Publons |
| F1000Prime recommendation | F1000Prime website |
| YouTube | youtube.com |
| Syllabus | Publicly available Syllabus of courses |
| Citations | Scopus and Web of Science |

### 1.1.6 X/Twitter and Facebook Data

For X/Twitter, Altmetric records identifiers of both the tweeter and tweet when a scientific output was mentioned on Twitter. Using the Twitter API (https://developer.twitter.com/en/docs/twitter-api/), we extracted the detailed information of all associated Twitter users (~10.2 million) and tweets (126 million) mentioning scientific work.

For individual Twitter users, we extracted the tweeter id, username (handle), display name, profile description, location description, follower count, following count, tweet count, creation date and verified account indicator. For each individual tweet, we extracted its identifier, text, retweet count, like count, reply count, quote count, created date, and referenced tweets if it's a retweet, reply tweet or quote tweet.

Given X/Twitter's dominant role among social media sources in mentioning scientific papers (covering 11.3 million articles in Altmetric), our main analysis focuses on X/Twitter. We also successfully replicate key findings when separately analyzing Facebook data, as it represents the second-largest social media source in Altmetric (covering 2.3 million articles).

**Table S2 Main Social Media Sources Tracked by Altmetric (accessed as of June 3, 2021).**

| Social Media Sources | Num. of Mentioned Papers |
|---|---|
| X/Twitter | 11,327,015 |
| Facebook | 2,271,555 |
| Reddit | 201,358 |
| YouTube | 175,541 |

### 1.1.7 Common Nicknames Dictionary

Based on the information collected from https://www.familiesunearthed.com/nicknames.htm, we constructed a common nickname to given name dictionary. In total, there are 764 pairs of mappings between nicknames and given names. We use these nicknames data when matching scientists to their tweets as describe below (see **S1.2.6**).

### 1.1.8 Coverage of News Outlets in United States

We verified the coverage of mainstream media outlets tracked by Altmetric against major news outlets listed by multiple sources. According to Nielsen, Hitwise, and comScore data, the Pew Research Center[1] identified 25 most popular news outlets in the United States, including four news aggregators such as Google News and The Examiner, as shown in **Table S3**. Of the 25 outlets, 21 out of 21 most popular news outlets are tracked by Altmetric. Among the remaining four news aggregators, Altmetric tracks both Google News and The Examiner (an aggregator of local news). Notably, Google News serves as the largest single driver of traffic to top news sites[2]. Thus, Altmetric tracks 100% of most popular specific media outlets and 92% when including online news aggregators.

**Table S3 Most Popular News Outlets Measured by Nielsen, comScore and Hitwise.**

| Newspaper Websites | | |
|---|---|---|
| The New York Times | Washington Post | USA Today |
| Wall Street Journal | LA Times | New York Daily News |
| New York Post | Boston Globe | San Francisco Chronicle |
| The Chicago Tribune | The British Daily Mail | |
| **Cable News Sites** | | |
| MSNBC | CNN | ABC News |
| Fox News | CBS News | BBC News |
| **Wire Service News Site** | | |
| Reuters | | |
| **Hybrid Online-only Sites** | | |
| Yahoo! News | AOL News | Huffington Post |
| **News Aggregators** | | |
| Google News | The Examiner | Topix |
| Bing News | | |

## 1.2 Methods
### 1.2.1 Matching Altmetric Database and OpenAlex Database

The Altmetric database provides high-scale data for mainstream media and social media mentions of specific scientific papers. However, Altmetric data lacks comprehensive bibliographic information for these scientific papers and, further, does not include papers that do not receive media attention. We therefore link records in the Altmetric to OpenAlex. By merging these datasets, we can not only study the characteristics of papers, authors, and institutions that are covered in media sources, we can also see what is not covered across the full landscape of science.

---
[1] https://www.pewresearch.org/journalism/2011/05/09/top-25/
[2] https://www.pewresearch.org/wp-content/uploads/sites/8/legacy/NIELSEN-STUDY-Copy.pdf

There are approximately 32.9 million scientific outputs recorded in Altmetric database across all output types, which include journal articles, conference papers, book chapters, datasets, books, dissertations, and others. In this study, we concentrated our focus on the journal articles recorded in Altmetric. Among these journal articles, 85% of them have DOIs (Digital Object Identifiers).

Our approach to linking journal articles within Altmetric to their corresponding records in the OpenAlex database was executed in two steps. First, we established links between journal articles with DOIs in Altmetric and their counterparts within the OpenAlex database. Second, for journal articles without DOIs or those not matched in the first step, we searched for them in OpenAlex database using bibliographic details, such as title, first author's name, journal name and publication year. Approximately 93.0% of journal articles documented in Altmetric were successfully matched with the corresponding publications in the OpenAlex database.

Finally, we focus on the 27,265 journals featured in the Scimago Journal Ranking that also have at least 1 paper matched with records in the Altmetric database. Focusing on SJR journals ensures that the papers come from the (large set of) journals that are prominent enough to be ranked and further ensures that, when looking at which papers are covered and which are not covered, we are tracking all works in a common set of journals across our datasets. Over the 2012-2019 timeframe, these journals include 20.9 million published articles and 14.2 million authors.

### 1.2.2 Gender Inference

Following (*4*), we use the NamSor API (https://namsor.app/) for name-to-gender inference. The NamSor API harnesses the predictive power of a Naive Bayes machine learning model, calibrated on 1.3 million unique names collected from different countries and regions. The model's predictive capabilities factor in linguistic and cultural nuances, enabling it to infer gender based on both first and last names, which contributes to better accuracy (*5*). NamSor API v2.0.26 was used to infer the gender of authors in our sample. Among the 14.2 million authors, 81.5% have full first names recorded in the OpenAlex database. Among those scientists with full first names, 58.7% of them are classified as men, while 41.3% of them are classified as women.

### 1.2.3 Novelty of Scientific Papers

Following prior research (*6*), we measure a paper's novelty by examining the combination of prior work in the paper's references. To compute novelty, we compare the observed frequency of journal pairs that appear within paper reference lists with a null model of the journal pair distribution created by randomized citation networks. Reference pairs that appear more than expected by chance are denoted as conventional and reference pairs that appear less than expected by chance are denoted as novel. Therefore, each paper has a distribution of z-scores for all journal pairs in its reference list. The novelty measure (*6*) is defined as the tenth percentile value of this distribution, where lower values indicate higher novelty. Details of the method can be found in pages 3 – 5 in the supplementary information of work (*6*).

The original measure introduced in the work of (*6*) follows a heavy-tail distribution. Therefore, we use a log transformation to convert the z-score to the form below. The new measure also improves readability, such that a higher score indicates greater novelty.

$$novelty = \begin{cases} -log_2(\text{z-score} + 1), & \text{z-score} > 0 \\ log_2(-\text{z-score} + 1), & \text{z-score} \leq 0 \end{cases} \quad (1)$$

For simplicity, we also define a binary variable *novel paper* as below:

$$novel\ paper = \begin{cases} 0, & \text{z-score} > 0 \\ 1, & \text{z-score} \leq 0 \end{cases} \quad (2)$$

The variable *novel paper* is used to indicate whether a paper is novel or not.

### 1.2.4 Impact of Scientific Papers

The OpenAlex database tracks citations between papers. We calculate the total number of citations received by paper $i$. We further normalize a paper $i$'s final citations, taking its ratio to the corresponding field and publication year mean. Denote this normalized ratio as $\hat{c}_i$. We first consider a binary indicator for whether the normalized citations are at the 95th percentile or above for that field and year, denoted $\widehat{c_i^{95}}$.

$$hit\ paper = \begin{cases} 0, & if\ \hat{c}_i < \widehat{c_i^{95}} \\ 1, & if\ \hat{c}_i \geq \widehat{c_i^{95}} \end{cases} \quad (3)$$

As an alternative measure, we also use $\hat{c}_i$ in a more continuous fashion. Given that $\hat{c}_i$ follows a heavy-tail distribution, we consider a log transform of $\hat{c}_i$,

$$impact = \log(\hat{c}_i + 1) \quad (4)$$

### 1.2.5 Team Composition

To investigate media coverage with regard to team demographics, we classify teams into the following categories: all women team (aw), all men team (am), mixed teams led by woman first author and woman last author (ww), mixed teams led by man first author and woman last author (mw), mixed teams led by woman first author and man last author (wm), and mixed teams led by man first author and man last author (mm). We further define the team composition variable $G_i \in \{aw, am, ww, mw, wm, mm\}$.

### 1.2.6 Matching Scientists on Twitter

There are 10.2 million Twitter accounts associated with tweets tracked by Altmetric (accessed June 3, 2021) that mentioned a specific scientific article. To match these Twitter users with authors documented in the OpenAlex database, our matching process primarily focuses on users who have self-promoted their own papers, following the methodology of Mongeon et al. (*7*). This approach operates under the assumption that scientists active on Twitter are also likely to disseminate their own work (*7*).

<u>Identifying Self-promoted Tweets</u>: To identify self-promoted tweets—where a scientist tweets about their own paper—we first extract the individual's name from their username (handle) and display name, following methodologies similar to those in (*7*). When deducing names, we remove any prefixes and suffixes commonly appended to usernames or display names, such as "Dr.," "Mr.," "Professor," "Prof.," "MD," "Jr.," and "PhD."

Unlike previous work (*7*), we also account for instances where tweeters use nicknames instead of their given names. For example, an author named "Benjamin Jones" may use "Ben Jones" on Twitter. To facilitate accurate name matching, we employed a nickname-to-given-name dictionary (see Section **S1.1.7**) for better matching performance.

The matching procedure first attempts to match complete first and complete last names, matching paper author information with the tweeter's username or display name. When we do not have the complete first name but rather the first name's initial (either in the publication database or from X/Twitter) we match on first initial and last name instead. Using this approach, we link 484,032 self-promoting tweeters with 499,440 authors in the OpenAlex database (similar with observations in the work of (*7*)). This results in a considerable number of one-to-many mappings between tweeters and OpenAlex authors. In the relatively rare cases where we do not have a one-to-one mapping, we further refine the matching procedure to identify the most plausible author-tweeter pair. We consider factors such as location data, incorporating both the author's institutional affiliation and the tweeter's stated location. Additionally, we consider author's research focus and the topics of papers tweeted by the tweeter—such as an economist tweeting about economics-related papers. Furthermore, we analyze collaborative and tweeting patterns, identifying cases where an author collaborates with a specific researcher, while a tweeter frequently shares papers authored by the same individual. This curation process results in a set of 476,890 unique tweeter-author pairs. Finally, among those identified unique tweeter-author pairs above, we focus on authors whose first publication years are between 1980 and 2020. With this additional period restriction, our main analysis in **Figure 4** considers 386k unique tweeter-author pairs.

### 1.2.7 Classification of Twitter Users

Altmetric classifies those who tweet scientific articles into four categories, and we follow a similar classification: **Scientist** - Twitter users who are likely to be scientists often have keywords like 'prof.', 'postdoc', 'asst prof', and 'scientist' in their profile descriptions; **Science Communicator** - Users in this group often have keywords like 'editor', 'correspondent', 'journalist', 'reporter', 'storyteller', and 'blogger' in their profile descriptions; **Healthcare Professional** - This category includes users with keywords related to healthcare, such as 'doctor', 'MD', 'public health', and 'nurse' in their profile descriptions; **Member of the Public** – other Twitter users. To further assess the identities in the members of the public category, we conduct a manual verification exercise (see Section **S1.2.8** below).

When considering our sample of 20.9 million papers published between 2012 and 2019, we can further categorize the source of initial tweets and retweets about specific papers. Collectively, scientists posted 4.9 million original tweets about these papers from 347.5 thousand Twitter accounts. These tweets were responded to in 11.2 million retweets. Among the retweeting accounts, 43.8% are classified as members of the public and 24.5% are classified as scientists. Separately, members of the public posted 1.8 million original tweets from 406.2 thousand Twitter accounts referencing specific scientific articles. These tweets were responded to in 1.3 million retweets. Among these retweeting accounts, 59.0% are classified as members of the public and 15.4% are classified as scientists.

In terms of follower counts among those who produced the initial tweets about specific scientific papers, the 347.5 thousand scientist accounts have an average of 1,983 followers and the 406.2 thousand accounts of members of the public average 2,535 followers. In addition to retweets,

Twitter users can respond actively by "likes" for a given tweet. The initial tweets referencing specific scientific articles received 24.9 million likes. Including retweets of these initial tweets, there were 36.0 million likes.

### 1.2.8 Members of the Public: Manual Verification

We randomly selected 150 Twitter accounts who are labeled as members of the public in our data. Of these 150 accounts, 114 are still active (as of Jan 25, 2025). For each of these, we investigated the account information and its tweets. We were able to identify occupation using profile descriptions for 38 Twitter accounts where this information was shared. We further were able to analyze tweet content for 107 of these Twitter accounts.

Reported occupations are shown below[3]. As can be seen in the table, these self-reported occupations do indeed appear to represent members of the public, with no evidence that any of them are research scientists.

| Manually Coded Job Info | ISCO Major Group (based on manual coded job info) |
|---|---|
| HR consultant | Professionals |
| Engineer | Professionals |
| Small business owner | Managers |
| Designer | Professionals |
| Farmer | Skilled Agricultural, Forestry, and Fishery Workers |
| Manager & industrial engineer | Managers |
| Artist | Professionals |
| Director of coaching & games development | Managers |
| Business professional in marketing | Professionals |
| Web developer | Professionals |
| Trader (old profile description) | Professionals |
| Lawyer | Professionals |
| Work in a tech company | Professionals |
| principal philanthropy specialist | Professionals |
| Manager | Managers |
| Artist | Professionals |
| "works with children" | Professionals |
| Teacher | Professionals |
| appears to be in a government position | Professionals |
| Writer | Professionals |
| Freelance writer | Professionals |
| Online safety expert | Professionals |
| An intern in a law-related position | Professionals |
| Drama graduate | Professionals |
| Web developer (based on old profile) | Professionals |
| MA student in Economics | Professionals |
| Actor | Professionals |
| Industrial employee | Professionals |

---

[3] The classification is based on International Standard Classification of Occupations (ISCO): https://ilostat.ilo.org/methods/concepts-and-definitions/classification-occupation/

| Criminologist | Professionals |
| --- | --- |
| Artist/writer | Professionals |
| Theatre educator | Professionals |
| Founder of "McDuck Enterprises" | Managers |
| Education related position | Professionals |
| Technical architect (based on old profile) | Professionals |
| A graduate student (based on old profile) | Professionals |
| Teacher | Professionals |
| Chef | Professionals |

Regarding tweet content, we conducted a thematic review using ChapGPT 4o. We used the Twitter API to download up to 100 tweets for each of 107 Twitter users. Then we asked GPT-4o (gpt-4o-2024-08-06) to summarize their themes in an automated manner. The prompt used for this analysis is: "You will be provided with a set of tweets/retweets from a Twitter user. Please summarize the main themes of these tweets." ChatGPT-4o classified the primary topics and their subtopics (given in parentheses) as follows:

- Personal and Lifestyle (Daily experiences & reflections; Food & Cooking; Travel & Adventure; Fitness & Exercise; Fashion & Beauty); – Some users focus on skincare, fashion trends, and beauty product recommendations.
- Media and Journalism (News & Journalism Integrity; Political Media Analysis; Social Media Influence; Independent Journalism & Citizen Reporting)
- Health and Wellness (Mental Health Advocacy; Pandemic & Public Health; Fitness & Nutrition; Medical System & Healthcare Access)
- Social Justice (Rights & Gender Identity; Feminism & Gender Equality; Racial & Ethnic Justice; Labor Rights & Workers' Movements; Disability Rights & Accessibility)
- Entertainment (Movies & TV Shows; Music & Celebrity News; Gaming & Esports; Book & Literature Discussions)
- Others (Politics & International Relations; Technology & AI; Environment & Climate Change; Humor & Memes)

Overall, the deeper dive and manual checking help clarify the nature of the "members of the public" category.

# 2 Regression Analysis

The results in **Figure 2** are further supported by fixed-effect ordinary least squares regressions. In this section, we discuss the details of our regression analyses.

## 2.1 Regression Analysis of Figure 2a.

Our results in **Figure 2a** are supported by fixed-effect ordinary least squares regressions as below.

$$y_i = \beta_m m_i + \beta_t t_i + \sum_t \beta_t T_{ti} + \sum_r \beta_r R_{ri} + \sum_d \beta_d D_{di} + \sum_e \beta_e E_{ei} + \sum_a \beta_a A_{ai} + \sum_h \beta_h H_{hi} \quad (5)$$
$$+ \sum_p \beta_p P_{pi} + \sum_q \beta_q Q_{qi} + \sum_f \beta_f F_{fi} + \sum_j \beta_j J_{ji} + \sum_g \beta_g G_{gi} + \epsilon_i$$

**Dependent Variable**: The dependent variable $y_i$ represents a paper's impact measured by the variable $impact$ defined in equation (4). An alternative measure is a binary variable $hit\ paper$ defined in equation (3), which indicates whether a paper is a hit paper being the top 5% home run papers gauged by citation.

**Predictors of Interest:** We use a binary variable $m_i$ to indicate whether a paper is mentioned by mainstream media outlets or not. A binary variable $t_i$ is used to indicate whether a paper is mentioned by Twitter or not. We consider regressions that control for these media mention indicators both separately and together.

**Control Variables:** We also include several other explanatory variables to control for other possible predictors of paper impact.
- $T_{ti}$: $T_{ti}$ indicates fixed effects that account for the size of a scientific team. We categorize a scientific team into 5 bins: $t = 1, t = 2, t = 3, t = 4,$ and $t = 5$ and $t \geq 6$. $T_{ti} = 1$ if the team size of a paper $i$ is in bin $t$ and $T_{ti} = 0$ otherwise.
- $R_{ri}$: $R_{ri}$ indicates fixed effects that account for the highest institution rank affiliated with a paper $i$. We categorize institution rank into 9 bins: [1, 10], [11, 20], [21, 40], [41, 80], [81, 160], [161, 320], [321, 640], [641, 1280], [1280, 1500] and no rank (no rank means that the institution is not recognized in U.S. News Ranking Database). $R_{ri} = 1$ if the highest institution rank affiliated with a paper $i$ is in bin $r$ and $R_{ri} = 0$ otherwise.
- $D_{di}$: $D_{di}$ indicates fixed effects that account for the career age of first author when a paper $i$ was published. We categorize the career age of first author into bins: [0, 5], [6, 10], [11, 15], [16, 20], [21, 25], [26, 30], [31, 35], [36, 40], [41, 45], [46, 50], [51, Inf]. $D_{di} = 1$ if the career age of first author is in a bin $d$ and $D_{di} = 0$ otherwise.
- $E_{Ei}$: $E_{ei}$ indicates fixed effects that account for the career age of last author when a paper $i$ was published, using the same bin definitions as for first authors above. $E_{ei} = 1$ if the career age of last author is in a bin $e$ and $E_{ei} = 0$ otherwise.
- $A_{ai}$: $A_{ai}$ indicates fixed effects that account for the average career age of a team, using the same bin definitions as for first authors above. $A_{ai} = 1$ if the average career age is in a bin $a$ and $A_{ai} = 0$ otherwise.
- $H_{hi}$: $H_{hi}$ indicates fixed effects that account for the impact of first author when a paper $i$ was published. We categorize the impact of first author into 15 exponential bins. $H_{hi} = 1$ if the impact of first author is in an exponential bin $h$ and $H_{hi} = 0$ otherwise.
- $P_{pi}$: $P_{pi}$ indicates fixed effects that account for the impact of last author when a paper $i$ was published. It has a similar setting as $H_{hi}$.

- $Q_{qi}$: $Q_{qi}$ indicates fixed effects that account for the average impact of authors when a paper $i$ was published. It has a similar setting as $H_{hi}$.
- $F_{fi}$: $F_{fi}$ indicates fixed effects that account for the field. Therefore, $F_{fi} = 1$ if a paper $i$ belongs to the field $f$ and $F_{fi} = 0$ otherwise.
- $J_{ji}$: $J_{ji}$ indicates fixed effects that account for the journal-year. For example, 'Science' and '2020' is one journal-year pair indicating all papers published by Science in the year of 2020. Therefore, $J_{ji} = 1$ if a paper $i$ belongs to the journal-year pair j and $J_{ji} = 0$ otherwise.
- $G_i$: $G_i$ indicates fixed effects that account for team demographics. The detailed definition can be found in Section **S1.2.5**.

The results of regression analyses are presented in **Table S4**. The results in models (1-3) confirm that there is a strong correlation between a paper's impact and its mainstream media attention and its Twitter attention after controlling for many other explanatory variables as listed above. The results in models (4-6) further confirm the results using a binary variable $hit\ paper$ to measure impact. The regressions show that papers covered by either mainstream media or Twitter tend to have substantially higher impact compared to those not covered, with the relationship being even stronger for mainstream media. These findings align with our observations in **Figure 2a**.

**Table S4 Relationship between media attention and paper impact**. In models (1), (2) and (3), a continuous variable is used to measure a paper's impact. In models (4), (5) and (6), a binary variable is used to indicate whether a paper is the top 5% home run paper gauged by citation. Both mainstream media mentions and Twitter mentions are strongly positively correlated with paper impact.

| Variable | Model (1) DV = $impact$ | Model (2) DV = $impact$ | Model (3) DV = $impact$ | Model (4) DV = $hit\ paper$ | Model (5) DV = $hit\ paper$ | Model (6) DV = $hit\ paper$ |
|---|---|---|---|---|---|---|
| Mainstream ($m_i$) | 0.26*** (0.0031) | | 0.21*** (0.002) | 0.21*** (0.0028) | | 0.19*** (0.0026) |
| Twitter ($t_i$) | | 0.14*** (0.0014) | 0.13*** (0.0013) | | 0.071*** (0.0011) | 0.057*** (0.00090) |
| Controls† | Y | Y | Y | Y | Y | Y |
| Observations | 17,592,230 | 17,592,230 | 17,592,230 | 17,592,230 | 17,592,230 | 17,592,230 |
| R Squared | 0.25 | 0.27 | 0.29 | 0.095 | 0.082 | 0.11 |

†The detail of control variables and fixed effects can be found in section **2.1**.
*, p < 0.05; **, p < 0.01; ***, p < 0.001.

### 2.1.1 Relationship between contemporary media attention and paper impact

We further explore the association between contemporary media attention and a paper's impact. Contemporary media attention is defined by mainstream media and Twitter mentions that occur within one year from the paper's publication date. This enables us to assess whether media attention can serve as a predictor of a paper's future impact.

**Table S5** shows that both mainstream media attention and Twitter attention offers early-stage prediction of a paper's future impact. First, in models (1-3), the results confirm a robust correlation between a paper's future impact and its contemporary mainstream media attention and Twitter attention. Second, in models (4-6), a binary variable, $hit\ paper$, is employed to determine whether a paper belongs to the top 5% of high-impact papers, as indicated by citation counts. Similar to the findings in **Table S4**, papers receiving contemporary mainstream media attention are nearly three times more likely to be home run papers than those mentioned in tweets (model (6)).

**Table S5 Relationship between contemporary media attention and paper impact**. In models (1), (2) and (3), a continuous variable is used to measure a paper's impact. In models (4), (5) and (6), a binary variable is used to indicate whether a paper is the top 5% home run paper gauged by citation. The results demonstrate that contemporary mainstream media mention and Twitter mention are positively predictive of a paper's future impact.

| Variable | Model (1) DV = $impact$ | Model (2) DV = $impact$ | Model (3) DV = $impact$ | Model (4) DV = $hit\ paper$ | Model (5) DV = $hit\ paper$ | Model (6) DV = $hit\ paper$ |
|---|---|---|---|---|---|---|
| Mainstream ($m_i$) | 0.22*** (0.0038) | | 0.18*** (0.0034) | 0.19*** (0.0037) | | 0.17*** (0.0035) |
| Twitter ($t_i$) | | 0.13*** (0.0015) | 0.12*** (0.0013) | | 0.067*** (0.0011) | 0.058*** (0.00096) |
| Controls† | Y | Y | Y | Y | Y | Y |
| Observations | 17,592,230 | 17,592,230 | 17,592,230 | 17,592,230 | 17,592,230 | 17,592,230 |
| R Squared | 0.24 | 0.26 | 0.27 | 0.082 | 0.080 | 0.092 |

†The detail of control variables and fixed effects can be found in section **2.1**.
*, $p < 0.05$; **, $p < 0.01$; ***, $p < 0.001$.

### 2.1.2  Relationship between retrospective media attention and paper impact

**Table S6 Relationship between retrospective media attention and paper impact**. In models (1), (2) and (3), a continuous variable is used to measure a paper's impact. In models (4), (5) and (6), a binary variable is used to indicate whether a paper is being the top 5% home run paper gauged by citation. The results demonstrate that retrospective mainstream media mention and Twitter mention are positively correlated with a paper's impact.

| Variable | Model (1) DV = $impact$ | Model (2) DV = $impact$ | Model (3) DV = $impact$ | Model (4) DV = $hit\ paper$ | Model (5) DV = $hit\ paper$ | Model (6) DV = $hit\ paper$ |
|---|---|---|---|---|---|---|
| Mainstream ($m_i$) | 0.37*** (0.0032) | | 0.29*** (0.0027) | 0.31*** (0.0028) | | 0.26*** (0.0023) |
| Twitter ($t_i$) | | 0.21*** (0.0018) | 0.18*** (0.0015) | | 0.15*** (0.0019) | 0.13*** (0.0015) |
| Controls† | Y | Y | Y | Y | Y | Y |
| Observations | 17,592,230 | 17,592,230 | 17,592,230 | 17,592,230 | 17,592,230 | 17,592,230 |
| R Squared | 0.25 | 0.26 | 0.28 | 0.093 | 0.095 | 0.11 |

†The detail of control variables and fixed effects can be found in section **2.1**.
*, $p < 0.05$; **, $p < 0.01$; ***, $p < 0.001$.

Additionally, we conducted an examination to evaluate the robustness of our findings from **Figure 2a** when considering only retrospective media attention. Retrospective media attention pertains to mentions in mainstream media and Twitter that occur after at least one year from the paper's publication date. The outcomes are detailed in **Table S6**, which confirms a substantial, positive correlation between retrospective media attention and paper impact.

### 2.1.3  Robustness Test with Self-promoted Tweets Removed

We further consider the findings excluding all self-promoted tweets (as outlined in Section **S1.2.6**). The results are detailed in **Table S7**. Models (1-2) confirm a substantial and positive correlation between Twitter attention and paper impact when self-promoted tweets are removed. Additionally, models (3-4) remain consistent when we use a binary variable ($hit\ paper$).

**Table S7 Relationship between Twitter attention and paper impact excluding self-promoted tweets**. In models (1) and (2), a continuous variable is used to measure a paper's impact. In models (3) and (4), the binary impact variable is used. The results demonstrate that Twitter attention is strongly positively correlated with a paper's impact when self-promoted tweets and scientists' tweets are removed.

| Variable | Model (1) DV = $impact$ | Model (2) DV = $impact$ | Model (3) DV = $hit\ paper$ | Model (4) DV = $hit\ paper$ |
|---|---|---|---|---|
| Mainstream ($m_i$) |  | 0.21*** (0.0027) |  | 0.19*** (0.0027) |
| Twitter ($t_i$) | 0.14*** (0.0014) | 0.13*** (0.0013) | 0.071*** (0.0011) | 0.058*** (0.00091) |
| Controls† | Y | Y | Y | Y |
| Observations | 17,592,230 | 17,592,230 | 17,592,230 | 17,592,230 |
| R Squared | 0.27 | 0.29 | 0.082 | 0.11 |

†The detail of control variables and fixed effects can be found in section **2.1**.
*, p < 0.05; **, p < 0.01; ***, p < 0.001.

## 2.2 Regression Analysis of Figure 2b.

Our results in **Figure 2b** are further analyzed using fixed-effect ordinary least squares regressions as below.

$$y_i = \beta_m m_i + \beta_t t_i + \sum_t \beta_t T_{ti} + \sum_r \beta_r R_{ri} + \sum_d \beta_d D_{di} + \sum_e \beta_e E_{ei} + \sum_a \beta_a A_{ai} \quad (6)$$
$$+ \sum_h \beta_h H_{hi} + \sum_p \beta_p P_{pi} + \sum_q \beta_q Q_{qi} + \sum_f \beta_f F_{fi} + \sum_j \beta_j J_{ji} + \sum_g \beta_g G_{gi} + \epsilon_i$$

**Dependent Variable**: The dependent variable $y_i$ indicates a paper's novelty measured by the variable $novelty$ defined in equation (1). An alternative measure is a binary variable $novel\ paper$ defined in equation (2), which indicates whether a paper is novel or not.

**Predictors of Interest:** We use a binary variable $m_i$ to indicate whether a paper is mentioned by mainstream media outlets or not. A binary variable $t_i$ is used to indicate whether a paper is mentioned by Twitter or not.

**Control Variables:** The detail of control variables can be found in **Section 2.1** above.

**Table S8 Relationship between media attention and paper novelty**. In models (1), (2) and (3), a continuous variable is used to measure a paper's novelty (see Equation (1)). In models (4), (5) and (6), a binary variable is used to indicate whether a paper is novel or not based on the definition in Equation (2). The results demonstrate that both mainstream media mention and Twitter mention are positively correlated with paper novelty.

| Variable | Model (1) DV = $novelty$ | Model (2) DV = $novelty$ | Model (3) DV = $novelty$ | Model (4) DV = novel $paper$ | Model (5) DV = novel $paper$ | Model (6) DV = novel $paper$ |
|---|---|---|---|---|---|---|
| Mainstream ($m_i$) | 0.68*** (0.015) |  | 0.61*** (0.015) | 0.12*** (0.0021) |  | 0.090*** (0.0018) |
| Twitter ($t_i$) |  | 0.29*** (0.011) | 0.25*** (0.011) |  | 0.10*** (0.0017) | 0.098*** (0.0017) |
| Controls† | Y | Y | Y | Y | Y | Y |
| Observations | 12,774,497 | 12,774,497 | 12,774,497 | 12,774,497 | 12,774,497 | 12,774,497 |
| R Squared | 0.070 | 0.070 | 0.071 | 0.11 | 0.12 | 0.12 |

†The detail of control variables and fixed effects can be found in section **2.1**.
*, p < 0.05; **, p < 0.01; ***, p < 0.001.

The results of the regression analysis can be found in **Table S8**. First, models (1-3) confirm a robust correlation between a paper's novelty and its mainstream media attention and Twitter attention. Second, models (4-6) present consistent results when we use a binary variable, *novel paper*, to determine whether a paper is novel or not. These results align with our previous observations regarding paper impact. Notably, papers covered by mainstream media and Twitter tend to be more novel in comparison to those not covered by these platforms. Furthermore, papers covered by mainstream media are also more novel compared to those covered by Twitter, as indicated in models (1)-(3). This finding corroborates our results presented in **Figure 2b**.

### 2.2.1 Relationship between contemporary media attention and paper novelty

In a similar fashion, we also investigated the relationship between contemporary media attention and paper novelty. The results presented in **Table S9** confirm that both contemporary mainstream media attention and Twitter attention exhibit a positive correlation with paper novelty.

**Table S9 Relationship between contemporary media attention and paper impact**. In models (1), (2) and (3), a continuous variable is used to measure a paper's impact. In model (4), (5) and (6), a binary variable is used to indicate whether a paper is being the top 5% home run paper gauged by citation. The results demonstrate that contemporary mainstream media mention and Twitter mention are positively predictive of a paper's future impact.

|  | Model (1) | Model (2) | Model (3) | Model (4) | Model (5) | Model (6) |
|---|---|---|---|---|---|---|
| Variable | DV = *novelty* | DV = *novelty* | DV = *novelty* | DV = novel *paper* | DV = novel *paper* | DV = novel *paper* |
| Mainstream ($m_i$) | 0.59*** |  | 0.53*** | 0.11*** |  | 0.078*** |
|  | (0.019) |  | (0.019) | (0.0026) |  | (0.0023) |
| Twitter ($t_i$) |  | 0.25*** | 0.23*** |  | 0.10*** | 0.096*** |
|  |  | (0.012) | (0.011) |  | (0.0018) | (0.0017) |
| Controls[†] | Y | Y | Y | Y | Y | Y |
| Observations | 12,774,497 | 12,774,497 | 12,774,497 | 12,774,497 | 12,774,497 | 12,774,497 |
| R Squared | 0.069 | 0.069 | 0.070 | 0.11 | 0.12 | 0.12 |

†The detail of control variables and fixed effects can be found in section **2.1**.
*, $p < 0.05$; **, $p < 0.01$; ***, $p < 0.001$.

### 2.2.2 Relationship between retrospective media attention and paper novelty

Additionally, we explored whether our findings from **Figure 2b** remain consistent when focusing solely on retrospective media attention. The results are presented in **Table S10**, providing similar findings to those presented in **Tables S8** and **S9**.

**Table S10 Relationship between retrospective media attention and paper novelty**. In models (1), (2) and (3), a continuous variable is used to measure novelty. In models (4) - (6), a binary variable is used to indicate whether a paper is novel or not. The results indicate that retrospective mainstream media mention and Twitter mention are positively correlated with a paper's novelty.

|  | Model (1) | Model (2) | Model (3) | Model (4) | Model (5) | Model (6) |
|---|---|---|---|---|---|---|
| Variable | DV = *novelty* | DV = *novelty* | DV = *novelty* | DV = novel *paper* | DV = novel *paper* | DV = novel *paper* |
| Mainstream ($m_i$) | 0.95*** |  | 0.76*** | 0.16*** |  | 0.12*** |
|  | (0.014) |  | (0.013) | (0.0019) |  | (0.0016) |
| Twitter ($t_i$) |  | 0.56*** | 0.49*** |  | 0.11*** | 0.10*** |
|  |  | (0.0082) | (0.0077) |  | (0.0014) | (0.0013) |
| Controls[†] | Y | Y | Y | Y | Y | Y |
| Observations | 12,774,497 | 12,774,497 | 12,774,497 | 12,774,497 | 12,774,497 | 12,774,497 |
| R Squared | 0.070 | 0.070 | 0.071 | 0.11 | 0.11 |  |

†The detail of control variables and fixed effects can be found in section **2.1**.
*, $p < 0.05$; **, $p < 0.01$; ***, $p < 0.001$.

### 2.2.3 Robustness Test with Self-promoted Tweets Removed

We further consider novelty when self-promoted tweets are excluded (as detailed in Section **S1.2.7**). The results, presented in **Table S11**, confirm that the novelty findings remain robust when removing self-promoted tweets.

**Table S11 Relationship between Twitter attention and paper novelty excluding self-promoted tweets**. In models (1) and (2), a continuous variable is used to measure a paper's novelty. In models (3) and (4), a binary variable is used to indicate whether a paper is novel or not. The results demonstrate that Twitter attention is still positively correlated with a paper's novelty when self-promoted tweets and scientists' tweets are removed.

| Variable | Model (1) DV = $novelty$ | Model (2) DV = $novelty$ | Model (3) DV = $novel\ paper$ | Model (4) DV = $novel\ paper$ |
|---|---|---|---|---|
| Mainstream ($m_i$) |  | 0.61*** (0.015) |  | 0.090*** (0.0018) |
| Twitter ($t_i$) | 0.28*** (0.012) | 0.24*** (0.011) | 0.10*** (0.0017) | 0.097*** (0.0017) |
| Controls[†] | Y | Y | Y | Y |
| Observations | 12,774,497 | 12,774,497 | 12,774,497 | 12,774,497 |
| R Squared | 0.070 | 0.071 | 0.12 | 0.12 |

†The detail of control variables and fixed effects can be found in section **2.1**.
*, $p < 0.05$; **, $p < 0.01$; ***, $p < 0.001$.

# 3 Impact and Novelty of Scientific Coverage

In this section, we conduct further robustness tests related to the impact and novelty findings in **Figure 2**.

## 3.1 Further Results on Impact

**Figure 2a** considers the citation distribution among papers that are mentioned in mainstream media and, separately, social media. Here we further investigate these citation distributions, but where we weight papers according to either (a) the number of mainstream media outlets and the number of Twitter users that mention the specific paper (**Figure S1a**) or (b) the relative circulation of the mainstream media outlets and number of Twitter followers (**Figure S1b**).

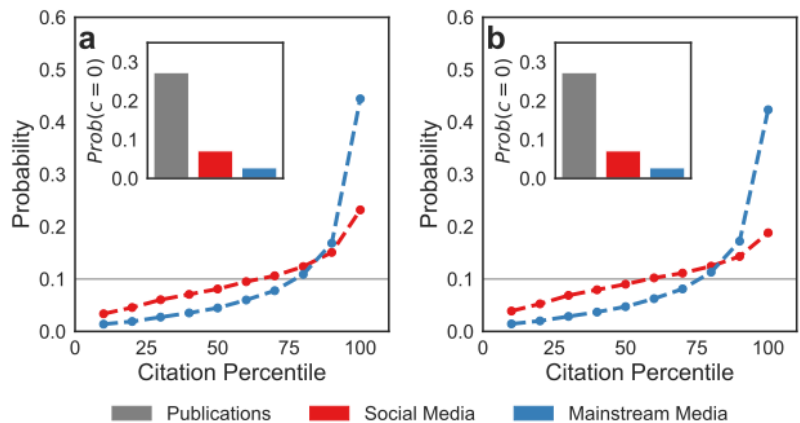

**Figure S1 Paper Impact Reported by the Media**. The x-axes indicate the citation percentile for papers with at least 1 citation ($c \geq 1$). The y-axes quantify the density. The inset figures report the percent of papers with zero citation among all journal articles (gray bar), the mainstream media reported articles (blue bar) and the scientist Twitter reported articles (red bar). (a) The paper citation percentile distributions are weighed by the number of mainstream media outlets and the number of Twitter users. (b) The paper citation percentile distributions are weighed by the mainstream media circulation and the number of Twitter followers.

This analysis reveals that, when weighting the citation distributions according to measures for audience size or the intensity of attention, the orientation on high impact work becomes even stronger. Mainstream media continues to show a greater impact orientation than social media. This suggests that highly impactful papers tend to garner larger audiences.

We further consider the analysis in **Figure 2a**, focusing on three subsamples:
(1) **Contemporary Mainstream Media/Twitter Mentions:** This sample focuses on mainstream media and Twitter mentions that occur within the same year as the publication of the paper.
(2) **Retrospective Mainstream Media/Twitter Mentions:** This sample focuses on mainstream media and Twitter mentions that appear more than one year after the paper's initial publication.
(3) **Excluding Self-Promoted Tweets:** This sample excludes Twitter mentions where authors tweeted about their own work.

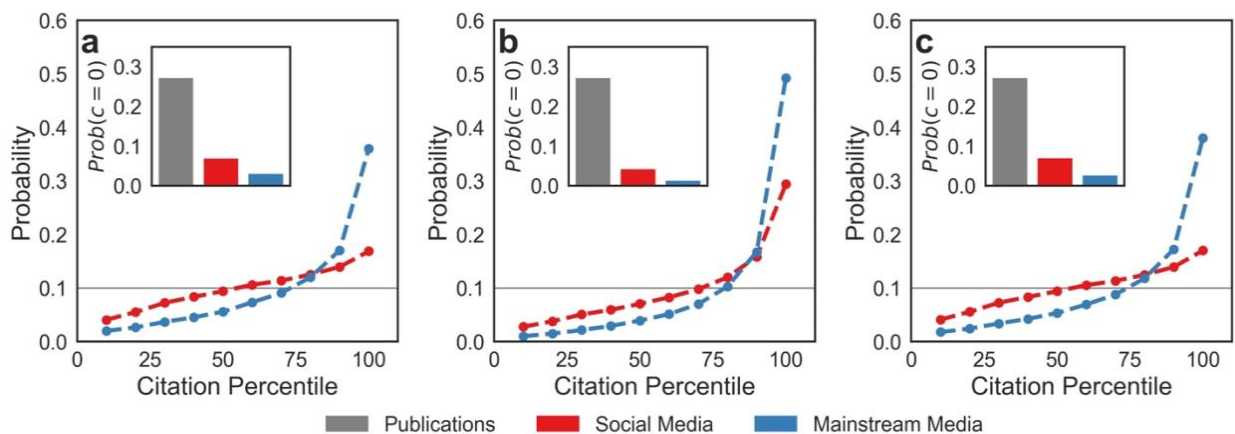

**Figure S2 Paper Impact Reported by the Media**. Gray/red/blue lines represent the probability density distributions of paper impact for all journal articles/the Twitter reported articles/the mainstream media reported articles. The inset figure reports the percent of papers with zero citation among all journal articles (gray bar), the Twitter reported articles (red bar), and the mainstream media reported articles (blue bar). (a) Paper citation percentile distribution when only contemporary mentions are considered. (b) Paper citation percentile distribution when only retrospective mentions are considered. (c) Paper citation percentile distribution when self-promoted tweets are excluded.

We observe in **Figure S2a** that both mainstream media and Twitter mentions present an early signal of a paper's impact. This observation aligns with the results obtained from our regression analysis, as detailed in Section **S2.1.1**. **Figure S2(b)** shows that papers covered by mainstream media, even in retrospect, maintain a higher level of impact compared to those mentioned on Twitter in a retrospective context. **Figure S2(c)** suggests that, when self-promoted tweets are excluded, papers mentioned by mainstream continue to show higher impact than those mentioned on Twitter.

### 3.2 Further Results on Novelty

**Figure 2b** considers the novelty distribution among papers that are mentioned in mainstream media and, separately, social media. As above for impact, we further investigate the novelty distributions, but where we weight papers according to either (a) the number of mainstream media outlets and the number of Twitter users that mention the specific paper (**Figure S3a**) or (b) the relative circulation of the mainstream media outlets and number of Twitter followers (**Figure S3b**).

Consistent with our previous observations in **Figure S1**, we find that papers covered by mainstream media and social media exhibit a higher level of novelty compared to all publications, and this relationship between novelty and media attention strengthens when we weight by the intensity of attention or audience size (**Figure S3**). As before, mainstream media continues to show a still greater orientation on novel work than social media.

In **Figure S4**, we further consider the analysis in **Figure 2b**, focusing on three subsamples: (a) contemporaneous media mentions; (b) retrospective media mentions; and (c) exclusion of self-promoted tweets. See **S3.1** above for further details on these subsamples. Overall, the findings for novelty are broadly similar across these different subsamples, as seen in **Figure S4** as well as in the regression analysis presented in **SS.2**.

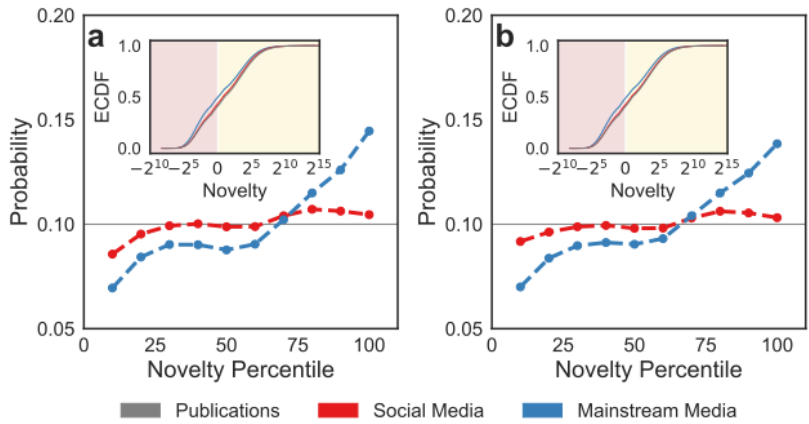

**Figure S3 Paper Novelty Reported by the Media.** Paper novelty measures the degree to which a paper combines past knowledge in a new way (*6*), with the measure converted to percentiles. Large percentile values indicate high levels of novelty. Gray/red/blue lines represent the probability density distributions of paper novelty for all journal articles/Twitter reported articles/mainstream media reported articles. The x-axes indicate the novelty percentile. The y-axes quantify the probability density. The inset figure represents the novelty z score's (*6*) cumulative density distribution. (a) The paper novelty percentile distribution is weighed by the number of mainstream media outlets and the number of Twitter users. (b) The paper citation percentile distribution is weighted by the number of mainstream media circulation and the number of Twitter followers.

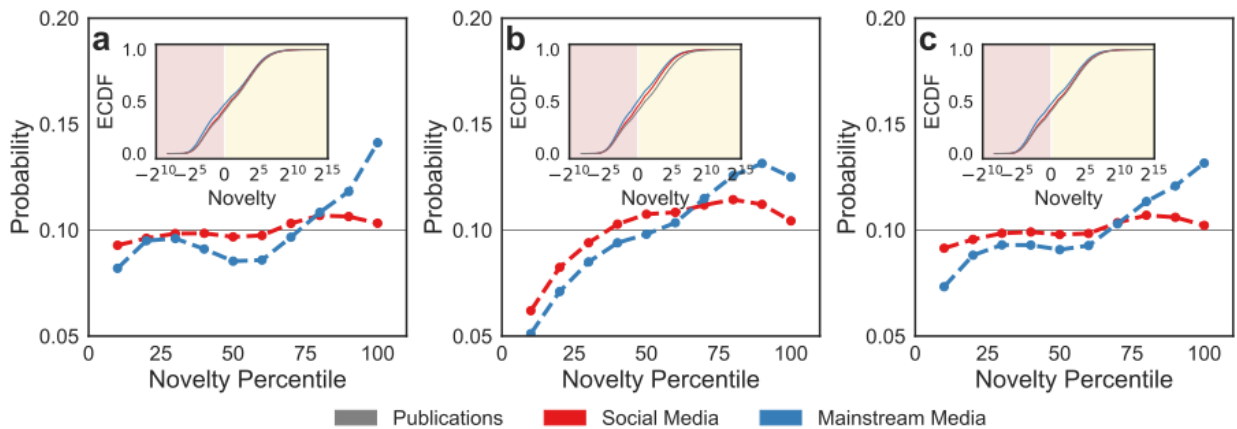

**Figure S4 Paper Novelty Reported by the Media.** (a) Paper novelty percentile distribution when only contemporary mentions are considered. Gray/red/blue lines represent the probability density distributions of paper novelty for all journal articles/the Twitter reported articles/the mainstream media reported articles. The inset figure represents the novelty z score's (*6*) cumulative density distribution. (b) Paper novelty percentile distribution when only retrospective mentions are considered. (c) Paper novelty percentile distribution when self-promoted tweets are excluded.

# 4 Heterogeneity of Scientific Coverage

**Figure 3** in the manuscript shows that social media presents more even scientific coverage in terms of field, institutional rank, journal rank and team demographic composition. In this section, we conduct several robustness tests.

## 4.1 Field Distribution of Reported Papers

In **Figure 3a**, our analysis focuses on the likelihood of a scientific paper being reported by mainstream media or mentioned on Twitter within each field of study. Our findings reveal that specific subject areas, such as Psychology, Biology, and Medicine, exhibit much higher rates of mainstream media coverage compared to other fields. By contrast, variation in coverage rates within Twitter are comparatively smaller, suggesting a more balanced distribution of attention across different scientific disciplines on the Twitter platform.

In **Figure S5**, we further consider the field distribution findings of Figure 3a, now focusing on three subsamples: (a) contemporaneous media mentions; (b) retrospective media mentions; and (c) exclusion of self-promoted tweets. See **S3.1** above for further details on these subsamples. Overall, the findings are broadly similar across these different subsamples.

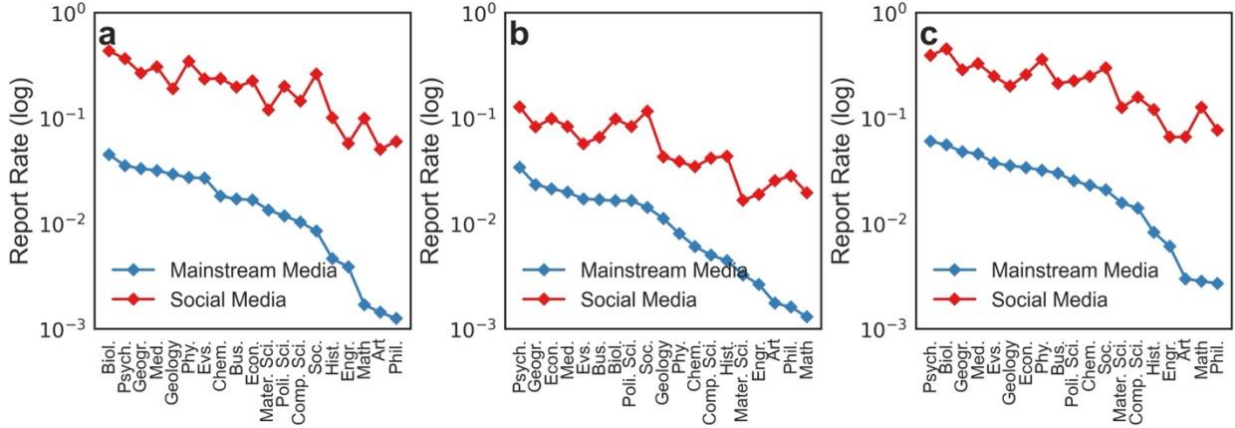

**Figure S5 Report Rate of Papers Covered by Mainstream Media and Twitter.** Red/blue lines represent the percentage of papers in each field that are reported by Twitter/mainstream media. The X-axis indicates each of the 19 fields. The Y-axis indicates the report rate. Fields have more even chances to be reported by Twitter than by mainstream media. (a) Report rate of papers covered by mainstream media and Twitter contemporarily. (b) Report rate of papers covered by mainstream media and Twitter retrospectively. (c) Report rate of papers covered by mainstream media and Twitter when self-promoted tweets are excluded.

### 4.1.1 Fixed Effect of Field in Predicting the Chance of Media Attention

To further investigate the observation that the distribution of scientific coverage across fields is flatter on Twitter, we run fixed-effect ordinary least squares regressions predicting the chance of being mentioned by mainstream media and Twitter respectively.

$$y_i = impact_i + novelty_i + \sum_t \beta_t T_{ti} + \sum_r \beta_r R_{ri} + \sum_d \beta_d D_{di} + \sum_e \beta_e E_{ei} + \sum_a \beta_a A_{ai} \qquad (7)$$
$$+ \sum_h \beta_h H_{hi} + \sum_p \beta_p P_{pi} + \sum_q \beta_q Q_{qi} + \sum_f \beta_f F_{fi} + \sum_j \beta_j J_{ji} + \sum_g \beta_g G_{gi} + \epsilon_i$$

In one regression, the dependent variable, $y_i$, is the number of mainstream media mentions. In a second regression, $y_i$, is the number of Twitter mentions. Descriptions of variables can be found in section **S2.1**.

We calculate the predicted margins for each field fixed effect in the regressions, which are further scaled by the mean of dependent variable in regression (7).

Additional regression analyses controlling for individual fixed effects yield results consistent with those presented in **Figures S6–S9**. Due to space constraints and redundancy, these results are not included here but are available upon request.

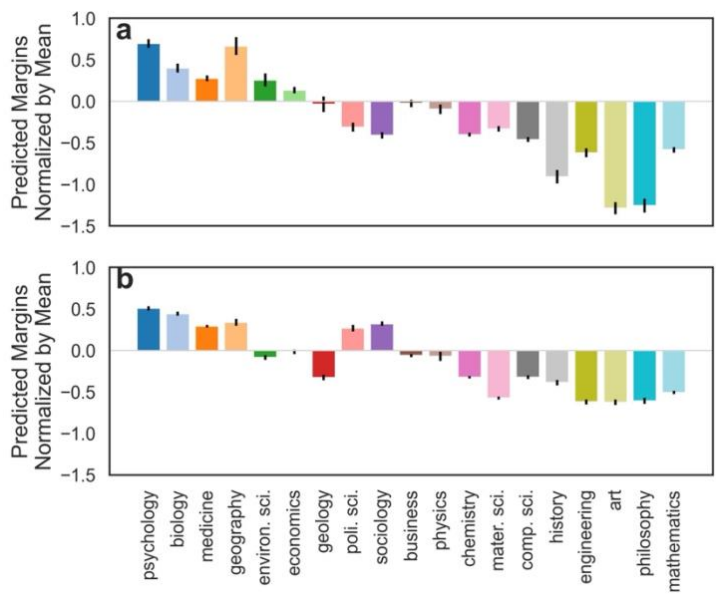

**Figure S6 Field Reporting Rates in Mainstream Media and Twitter**. (a) Field reporting rates in mainstream media, net of regression controls. (b) Field reporting rates in Twitter, net of regression controls. The X-axis indicates each field. The Y-axis quantifies margins of fixed effect for 19 scientific fields.

In **Figure S6**, the 19 fields are organized in the same order as **Figure 3a**, thereby facilitating a direct comparison. Net of regression controls, the findings are broadly similar. While there are shifts for some fields, the central insight remains that variations in reporting rates across fields are less pronounced in Twitter than in mainstream media. We further consider field differences for subsamples of the data: contemporary media mentions (**Figure S7**); retrospective media mentions (**Figure S8**); and when self-promoted tweets are excluded (**Figure S9**). The findings for these subsamples are all broadly consistent with the analysis of the full sample.

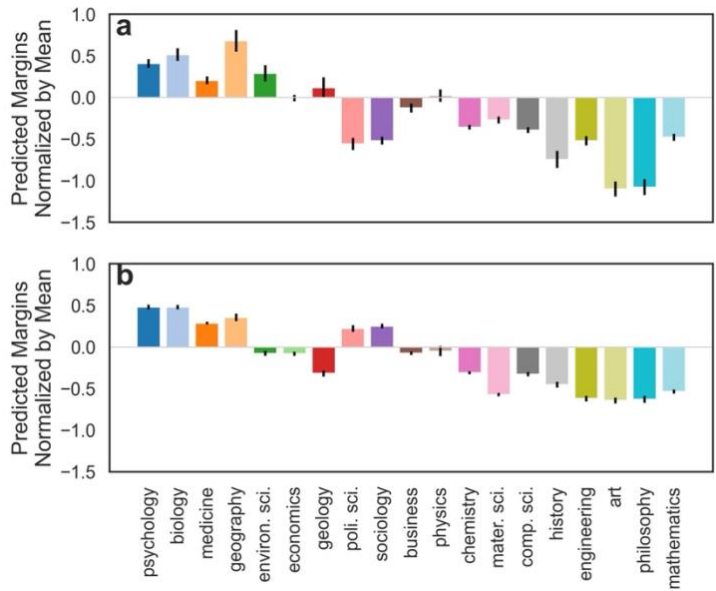

**Figure S7 Field Reporting Rates in Mainstream Media and Twitter, Contemporary Mentions Only**. (a) Field reporting rates in mainstream media, net of regression controls. (b) Field reporting rates in Twitter, net of regression controls. The X-axis indicates each field. The Y-axis quantifies margins of fixed effect for 19 scientific fields.

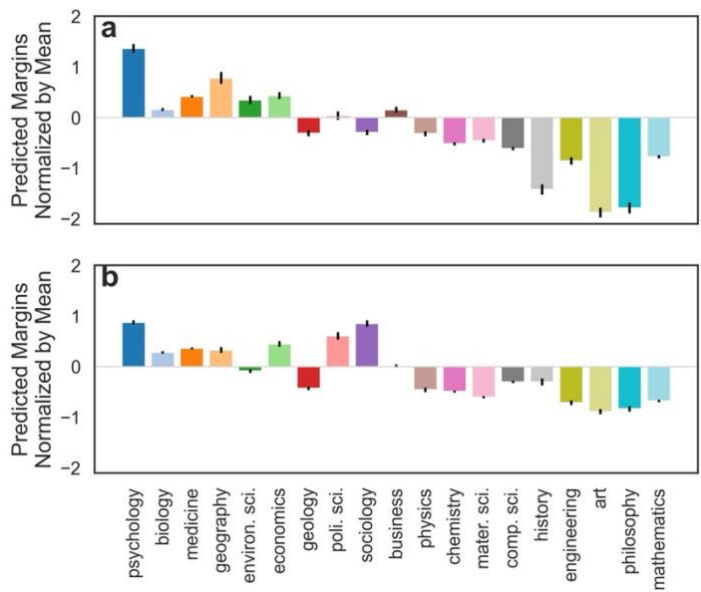

**Figure S8 Field Reporting Rates in Mainstream Media and Twitter, Retrospective Mentions Only**. (a) Field reporting rates in mainstream media, net of regression controls. (b) Field reporting rates in Twitter, net of regression controls. The X-axis indicates each field. The Y-axis quantifies margins of fixed effect for 19 scientific fields.

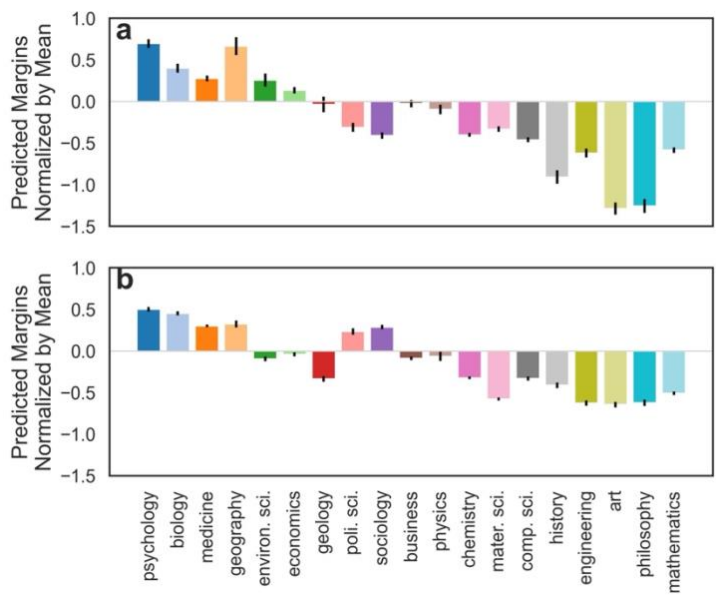

**Figure S9 Field Reporting Rates in Mainstream Media and Twitter, Excluding Self-Promoting Tweets**. (a) Field reporting rates in mainstream media, net of regression controls. (b) Field reporting rates in Twitter, net of regression controls. The X-axis indicates each field. The Y-axis quantifies margins of fixed effect for 19 scientific fields.

## 4.2 Institutional Rank Distribution of Reported Papers

**Figure 3b** and **3c** show that both mainstream media and Twitter coverage strongly emphasize papers from highly-ranked institutions. To test the robustness of these findings, we first consider whether they prevail when examining contemporaneous media mentions, retrospective media mentions, and when excluding self-promoted tweets. The results, presented in **Figure S10** and **Figure S11**, prove similar as when studying the full sample. One distinction is that Twitter is slightly more biased towards elite institutions when consider retrospective paper mentions (**Fig. S11b**).

We next consider reporting rates across institutions while accounting for the audience size or intensity of media mentions. In **Figure S12**, we present the cumulative distributions weighted by the number of mainstream media outlets and the number of Twitter users (**Figure S12a**), and with the cumulative distributions weighed by the circulation of mainstream media outlets and the followers' number of Twitter users (**Figure S12b**).

The results in **Figure S12** continue to show that both mainstream media and social media mentions orient especially on higher-ranked institutions. However, when weighting by the number of mentions or audience size, we see that this orientation on elite institutions is even more intense. Further, while social media continues to be in an intermediate state between mainstream media and the underlying citation distribution across institutions, social media is now closer to mainstream media in its orientation (compare **Figure S12** with **Figure 3c**).

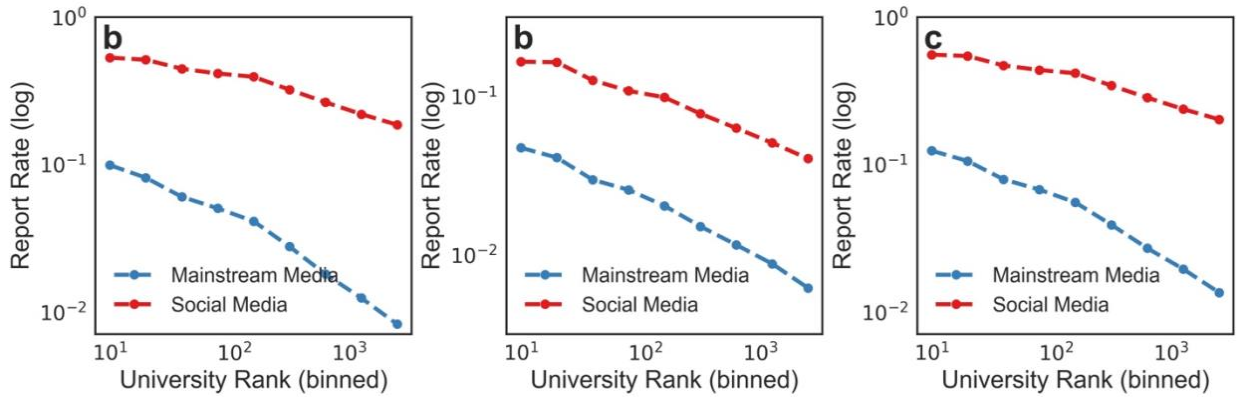

**Figure S10 Report Rate of Papers Covered by Mainstream Media and Twitter**. Red/blue lines indicate the report rates in Twitter/mainstream media. The figure presents (a) the report rate of papers covered by mainstream media and Twitter contemporaneously; (b) the report rate of papers covered by mainstream media and Twitter retrospectively; (c) the report rate of papers covered by mainstream media and Twitter where self-promoted tweets are removed.

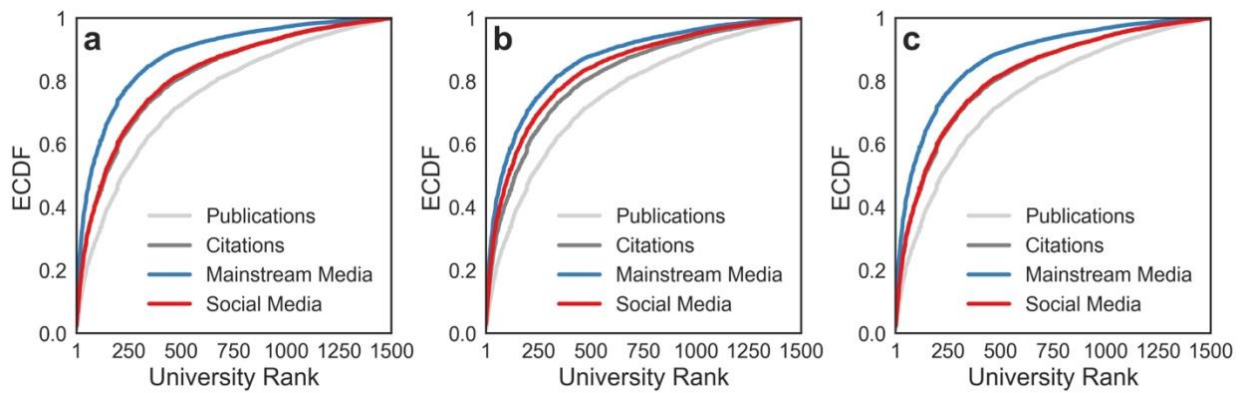

**Figure S11 Cumulative Distribution Functions by University Rank for Three Subsamples**. The empirical cumulative distributions of the rank of institutions, based on US News Best Global Universities ranking. The smaller the magnitude of the rank, the more highly ranked the institution. Light grey/grey/red/blue lines represent the ECDF of the institution rank of all publications/all citations/papers reported by Twitter/papers reported by mainstream media. (a) The ECDF for contemporary media mentions. (b) The ECDF for retrospective media mentions. (c) The ECDF where self-promoted tweets are removed.

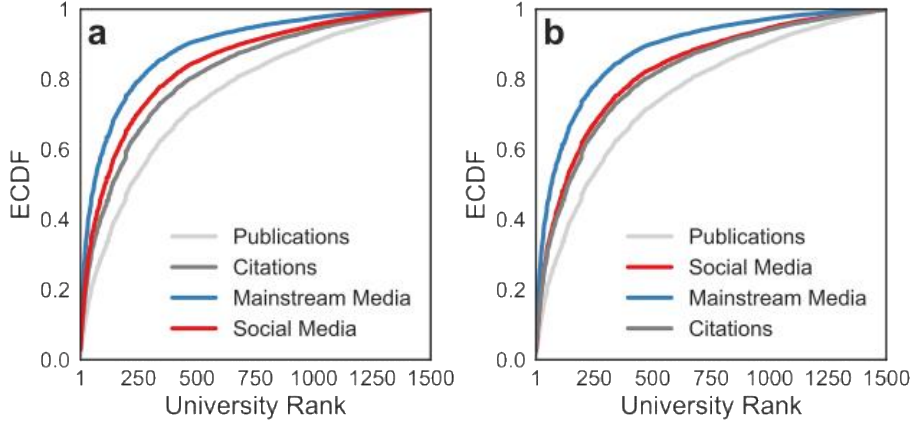

**Figure S12 Cumulative Distribution Functions by University Rank, with Attention Weights.** Light grey/grey/red/blue lines represent the ECDF of the institution rank of all publications/all citations/papers reported by Twitter/papers reported by mainstream media. (a) Paper mentions weighted by the number of mainstream media mentions and the number of Twitter user mentions. (b) Paper mentions weighted by the circulation of mainstream media outlets and the number of followers for the tweet.

### 4.2.1 Fixed Effect of Institutional Rank in Predicting the Chance of Media Attention

We further investigate the institutional relationships with media mentions using fixed-effect ordinary least squares regressions.

$$y_i = impact + novelty + \sum_t \beta_t T_{ti} + \sum_r \beta_r R_{ri} + \sum_d \beta_d D_{di} + \sum_e \beta_e E_{ei} + \sum_a \beta_a A_{ai} + \sum_h \beta_h H_{hi} \quad (8)$$
$$+ \sum_p \beta_p P_{pi} + \sum_q \beta_q Q_{qi} + \sum_f \beta_f F_{fi} + \sum_j \beta_j J_{ji} + \sum_g \beta_g G_{gi} + \epsilon_i$$

In one regression, the dependent variable, $y_i$, is the number of mainstream media mentions. In a second regression, $y_i$, is the number of Twitter mentions. Descriptions of variables can be found in section **S2.1**.

We compute the predicted margins of institutional rank fixed effects (see section **S2.1** for definition), which are further scaled by the mean of dependent variable in regression (8). The margins of institutional rank are presented visually in **Figure S13**. The results align with our observations in **Figure 3b** and **3c**. **Figure S13** demonstrates that mainstream media and social media disproportionately report on papers from top universities. At the same time, the gap in normalized margins in mainstream media coverage is approximately twice of that observed on Twitter, highlighting Twitter's role in evening the visibility of papers from institutions with varying degrees of prestige.

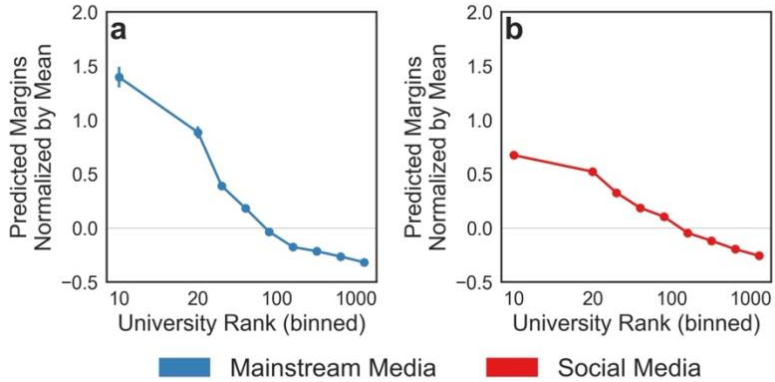

**Figure S13 Institutional Rank as a Predictor of Media Mentions**. (a) Margins of institutional rank in predicting mainstream media mentions. The X-axis indicates 9 exponential bins of institutional rank (see Section **S2.1**). Smaller values indicate higher ranks. The Y-axis quantifies the margins of each institutional rank level. (b) Margins of institutional rank in predicting the chance of being reported by Twitter.

We further conduct the regression analysis for three different subsamples of media mentions: contemporaneous mentions (**Figure S14**); retrospective mentions (**Figure S15**), and mentions excluding self-promoting tweets (**Figure S16**). The results are all broadly similar to the findings when studying the full sample. One distinction is that there is more convergence between mainstream media mentions and social media mentions when looking retrospectively at papers. This convergence is largely because mainstream media is somewhat less biased to the most elite institutions when looking retrospectively. One interpretation is that, with contemporaneous papers, journalists use the prestige of the institution as an information cue, while with the passage of time important papers may rely less on such institutional cues in garnering media attention.

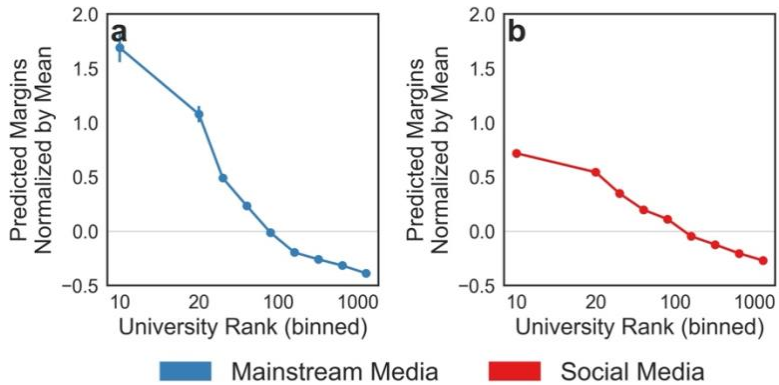

**Figure S14 Institutional Rank as a Predictor of Contemporaneous Media Mentions**. (a) Margins of institutional rank in predicting contemporaneous mainstream media mentions. (b) Margins of institutional rank in predicting contemporaneously Twitter mentions.

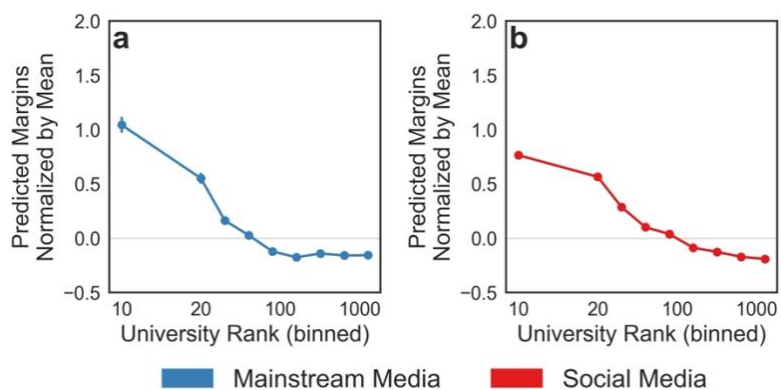

**Figure S15 Institutional Rank as a Predictor of Retrospective Media Mentions**. (a) Margins of institutional rank in predicting contemporaneous mainstream media mentions. (b) Margins of institutional rank in predicting contemporaneously Twitter mentions.

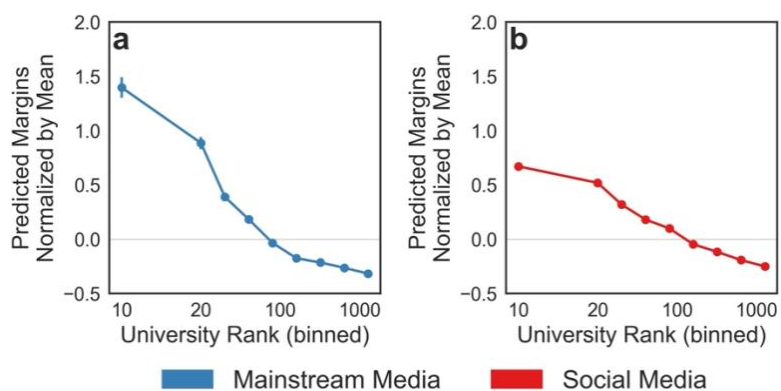

**Figure S16 Institutional Rank as a Predictor of Media Mentions, Self-promoted Tweets Removed**. (a) Margins of institutional rank in predicting contemporaneous mainstream media mentions. (b) Margins of institutional rank in predicting contemporaneously Twitter mentions, with self-promoted tweets removed.

Additional regression analyses controlling for individual fixed effects yield results consistent with those presented in **Figures S13–S16**. Due to space constraints and redundancy, these results are not included here but are available upon request.

### 4.3 Journal Rank Distribution of Reported Papers

We continue with similar robustness analysis, examining the findings for journal rank in **Figure 3d**. **Figure S17** reconsiders the main results for three different subsamples of media mentions: contemporaneous mentions (**Figure S17a**); retrospective mentions (**Figure S17b**), and mentions excluding self-promoting tweets (**Figure S17c**). In these subsamples, Twitter demonstrates a more balanced distribution of attention across various journals compared to mainstream media. As with institutional rank, when looking at retrospective mentions, the relationship between journal rank and media mentions has a more similar slope, comparing mainstream media and Twitter. One interpretation may be that, with contemporaneous papers, journalists use the journal prestige as an information cue, while retrospectively, with the passage of time, important papers may rely less on such journal cues in garnering media attention.

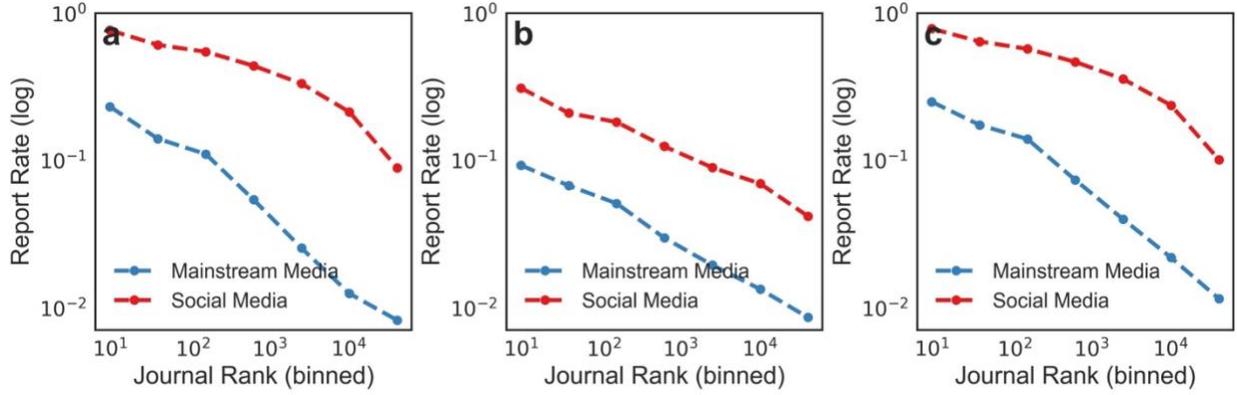

**Figure S17 Report Rate of Journals Covered by Mainstream Media and Twitter**. The X-axis indicates journals (exponentially binned Scimago journal rank). The Y-axis indicates the report rate in each rank level. Red/blue lines indicate the report rates of journal articles in Twitter/ mainstream media. The figures consider three subsamples of media mentions: (a) contemporaneous mentions. (b) retrospective mentions. (c) mentions excluding self-promoted tweets.

### 4.3.1 Fixed Effect of Journal Rank in Predicting Media Attention

To delve deeper into the observation highlighted above, we conducted two fixed-effect ordinary least squares (OLS) regressions to predict the likelihood of a paper being mentioned by mainstream media and Twitter, respectively.

$$y_i = impact_i + novelty_i + \sum_t \beta_t T_{ti} + \sum_r \beta_r R_{ri} + \sum_d \beta_d D_{di} + \sum_e \beta_e E_{ei} + \sum_a \beta_a A_{ai} \qquad (9)$$
$$+ \sum_h \beta_h H_{hi} + \sum_p \beta_p P_{pi} + \sum_q \beta_q Q_{qi} + \sum_f \beta_f F_{fi} + \sum_j \beta_j J_{ji} + \sum_g \beta_g G_{gi} + \epsilon_i$$

The description of dependent variables and control variables can be found in section **S2.1**. In this section, we categorize journals into 10 exponential bins based on their Scimago journal ranks: [1, 50], [51, 100], [101, 200], [201, 400], [401, 800], [801, 1600], [1601, 3200], [3201, 6400], [6401, 12800] and [12801, ∞]. We then compute the predicted margins of journal rank fixed effects.

**Figure S18** visualizes the predicted margins, which are further scaled by the mean of dependent variable in regression (9). Notably, papers published in the top 50 journals are almost seven times more likely to be covered by mainstream media than the population average. This is a significant contrast to Twitter, where papers from top 50 journals are only twice as likely to be mentioned compared to the population average. These regression findings continue to highlight that Twitter presents a more even distribution of journals compared to mainstream media outlets.

In **Figures S19-S21**, we consider further regressions for the subsets of the data: when studying contemporaneous media mentions (**Figure S19**), retrospective media mentions (**Figure S20**), and when excluding self-promoting tweets (**Figure S21**). The results are broadly similar when looking at these subsamples. As in analyses above, a distinctive difference is that mainstream media puts less emphasis on the most elite journals when looking at papers retrospectively, again potentially consistent with information cues as discussed above.

We also conducted additional regression analyses controlling for individual fixed effects, which yielded results consistent with those presented in **Figures S18–S21**. Due to space constraints, we do not include them here, but they are available upon request.

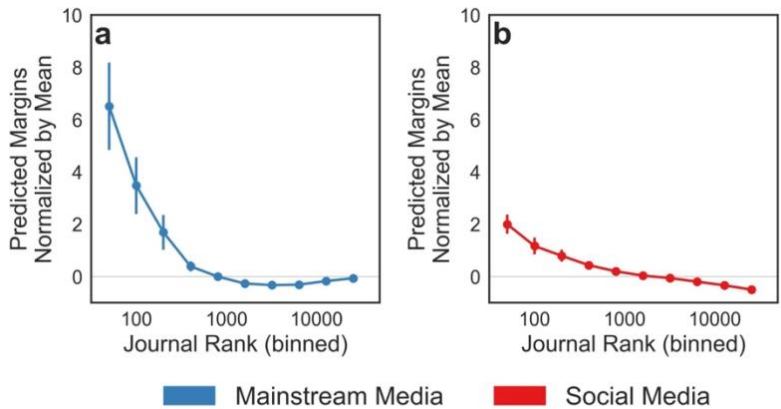

**Figure S18 Journal Rank as a Predictor of Media Mentions**. (a) Margins of journal rank in predicting mainstream media mentions. The X-axis indicates 10 exponential bins of journal rank. Smaller values indicate better ranks. The Y-axis quantifies the margins of each journal rank level. (b) Margins of journal rank in predicting Twitter mentions.

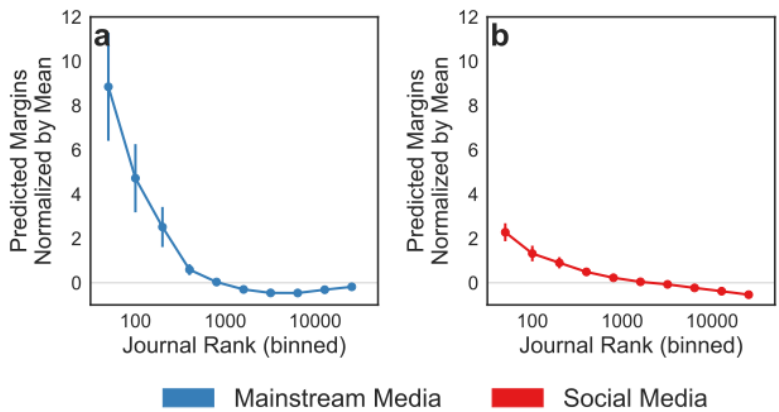

**Figure S19 Journal Rank as a Predictor of Contemporaneous Media Mentions**. (a) Margins of journal rank in predicting contemporaneous mainstream media mentions. (b) Margins of journal rank in predicting contemporaneous Twitter mentions.

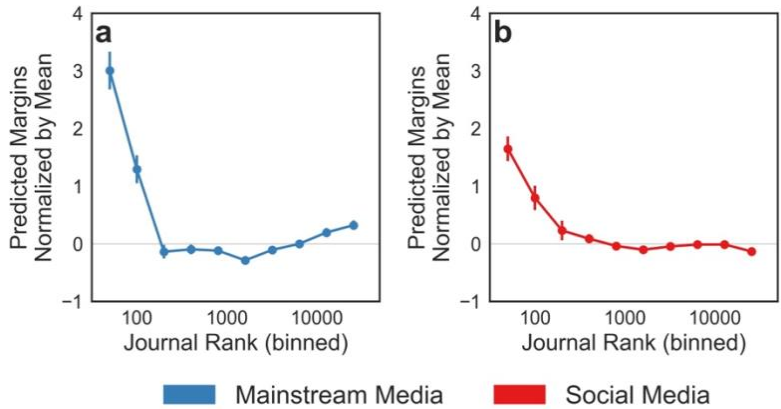

**Figure S20 Journal Rank as a Predictor of Retrospective Media Mentions**. (a) Margins of journal rank in predicting retrospective mainstream media mentions. (b) Margins of journal rank in predicting retrospective Twitter mentions.

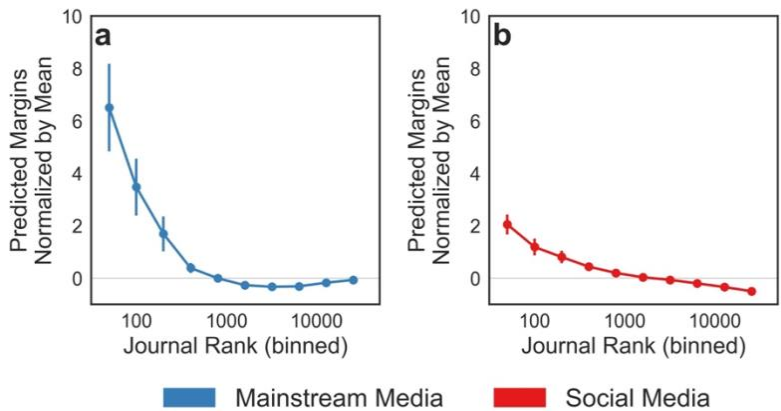

**Figure S21 Journal Rank as a Predictor of Media Mentions, Self-Promoting Tweets Excluded**. (a) Margins of journal rank in predicting mainstream media mentions. (b) Margins of journal rank in predicting Twitter mentions where self-promoted tweets are removed.

# 5 Team Composition

**Figures 3e** and **3f** consider the representation of teams with varied demographic compositions. A key observation is the reduced disparity between all-women teams (aw) and all-men teams (am) on Twitter compared to mainstream media. This finding shows that Twitter more evenly highlights the work of scientific teams, regardless of the demographic composition.

Here, we conduct several robustness tests for the findings in **Figure 3e** and **3f**. We examine whether our main finding holds when (1) only contemporaneous media mentions are included; (2) only retrospective media mentions are included; and (3) self-promoted tweets are excluded. With all controls, the fixed-effect ordinary least squares regressions take the form:

$$y_i = \sum_g \beta_g G_{gi} + impact_i + novelty_i + \sum_t \beta_t T_{ti} + \sum_r \beta_r R_{ri} + \sum_d \beta_d D_{di} + \sum_e \beta_e E_{ei} + \sum_a \beta_a A_{ai} \quad (10)$$
$$+ \sum_h \beta_h H_{hi} + \sum_p \beta_p P_{pi} + \sum_q \beta_q Q_{qi} + \sum_f \beta_f F_{fi} + \sum_j \beta_j J_{ji} + \epsilon_i$$

**Dependent Variable**: The dependent variable $y_i$ indicates either the number of mainstream media reports that a paper $i$ receives or the number of tweets that a paper $i$ receives. We further control for individual author fixed effects in this analysis.

**Predictors of Interest:** We use a categorical variable $G_i$ to indicate a paper's team composition (see **Section S1.2.6**).

**Control Variables:** The detail of control variables can be found in **Section 2.1** above.

We consider the team composition results, looking at subsamples of media mentions: contemporaneous mentions (**Figure S22**); retrospective mentions (**Figure S23**); mentions excluding self-promoting tweets (**Figure S24**). The results are broadly similar as to the whole sample, with the exception of retrospective Twitter mentions. Retrospective Twitter mentions look more like mainstream media mentions, with a greater under-representation of all-women teams and mixed teams led by women.

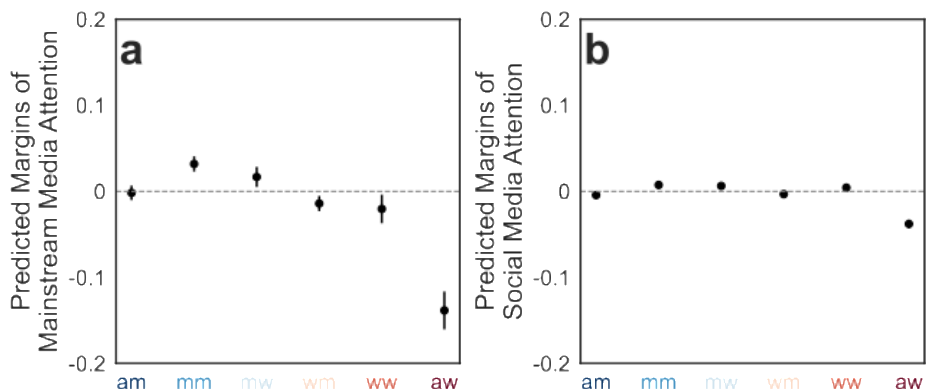

**Figure S22 Team Composition in Predicting Contemporaneous Media Mentions**. (a) Margins of team composition in predicting contemporaneous mainstream media mentions. (b) Margins of team composition in predicting contemporaneous Twitter mentions.

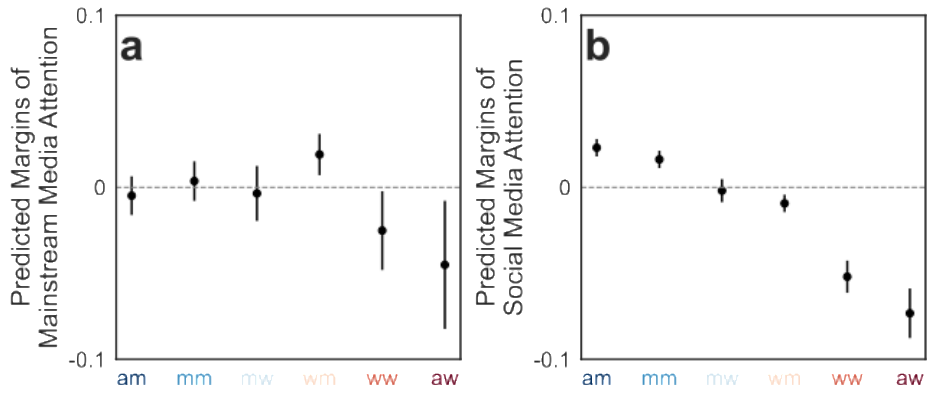

**Figure S23 Team Composition in Predicting Retrospective Media Mentions**. (a) Margins of team composition in predicting retrospective mainstream media mentions. (b) Margins of team composition in predicting retrospective Twitter mentions.

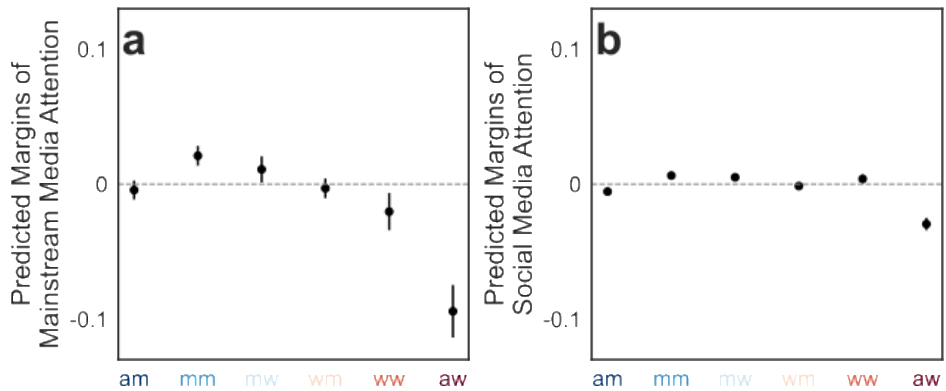

**Figure S24 Team Composition in Predicting Media Mentions, Self-promoted Tweets Removed.** (a) Margins of team composition in predicting mainstream media mentions. (b) Margins of team composition in predicting Twitter mentions where self-promoted tweets are removed.

# 6 Scientists on Twitter

In the following analysis, we focus on scientists, as described in Section **S1.2.8**, whose first publications appear from 1980 to 2020. We examine the dynamics of these scientists' engagement with Twitter and their interactions with the scientific community.

## 6.1 Matched Scientists on Twitter

**Figure S25** presents the distribution of matched scientists across various scientific fields. Sociology exhibits the highest percentage, with over 5.2% of scientists matched to a Twitter account. In contrast, engineering has the lowest proportion, with approximately 0.19% of scientists matched.

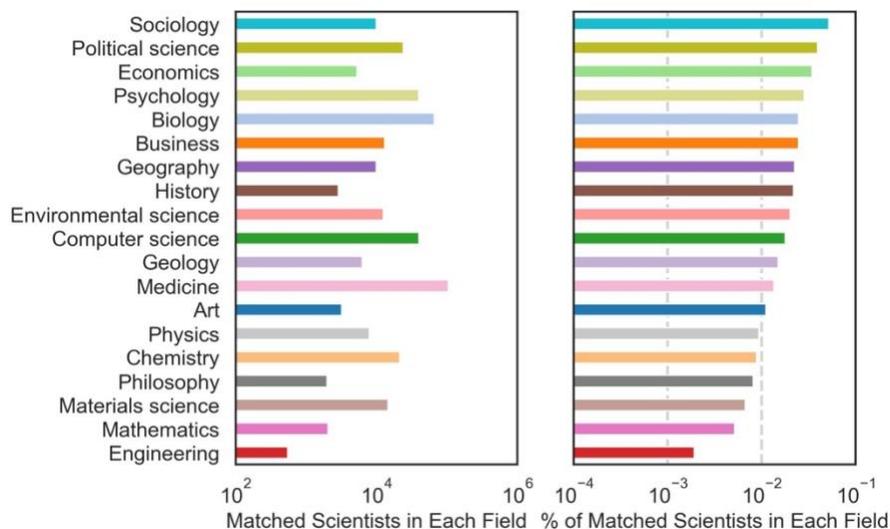

**Figure S25 Distribution of Matched Scientists over 19 Scientific Fields.** (Left) graphs the number of matched scientists in different scientific fields; (Right) graphs the proportion of matched scientists in different fields. The 19 scientific fields are sorted by the proportion of matched scientists from top to bottom. Medicine has the highest number of matched scientists, while engineering has the smallest number of matched scientists. Sociology has the highest proportion of matched scientists, while engineering has the lowest proportion of matched scientists.

We further show that, compared to the broader public on Twitter, matched scientists have similar characteristics in terms of follower counts and years active. **Figure S26** examines the distribution of the number of followers. We compare the follower count distribution for matched scientists with the follower count the broader public on Twitter. We see that these distributions are very similar. **Figure S27** considers the number of years individuals have been active on Twitter, comparing matched scientists with the broader universe of accounts. We again find that the distributions are very similar.

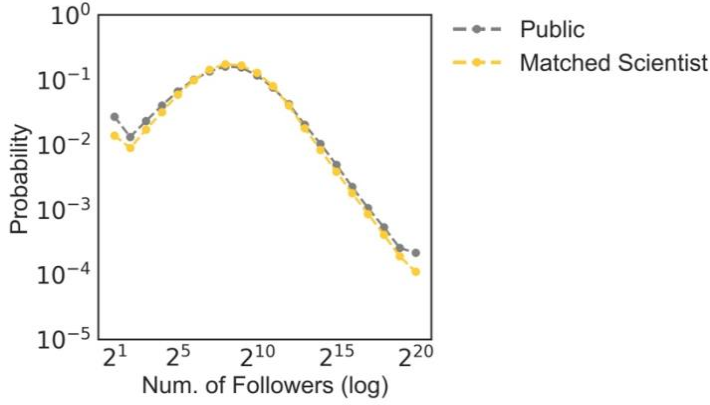

**Figure S26 Number of Followers Distributions.** Gray/Yellow lines indicate the number of followers distributions for non-scientists/matched scientists with Twitter accounts.

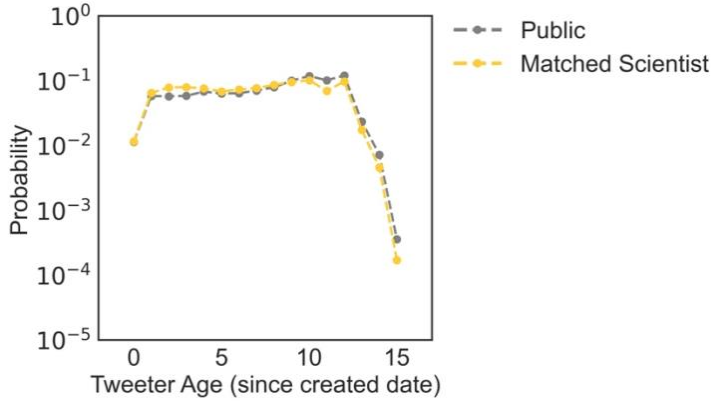

**Figure S27 Tweeter Age Distributions**. Gray/Yellow lines indicate the Tweeter age distributions for non-scientists/matched scientists with Twitter accounts.

### 6.2 Scientists' Citations and Number of Followers

**Figure 4a** considers the relationship between citation counts and follower counts for matched scientists. We further consider this relationship using a fixed-effect ordinary least squares regression as below.

$$y_i = \beta_c c_i + \sum_p \beta_p P_{pi} + \sum_r \beta_r R_{ri} + \sum_a \beta_a A_{ai} + \sum_f \beta_f F_{fi} + \sum_l \beta_l L_{li} + \epsilon_i \qquad (11)$$

**Dependent Variable**: The dependent variable $y_i$ measures an author $i$'s normalized number of followers by field. As a robustness check, we also use an author $i$'s normalized number of followers by field-year as an alternative dependent variable.

**Predictors of Interest:** The independent variable $c_i$ indicates an author $i$'s normalized citations by field. As a robustness check, we also use an author $i$'s normalized citations by field-year as an alternative dependent variable.

**Control Variables:** We also include several other explanatory variables to control for other possible predictors.

- $P_{pi}$: $P_{pi}$ indicates fixed effects that account for an author $i$'s total number of publications. We categorize the total number of publications into 9 bins: [0,0], [1,2], [3,4], [5,8], [9, 16], [17, 32], [33, 64], [65, 128], [129, Inf]. $P_{pi} = 1$ if the total number of publications is in a bin $p$ and $P_{pi} = 0$ otherwise.
- $R_{ri}$: $R_{ri}$ indicates fixed effects that account for the highest institution rank affiliated with an author $i$. We categorize institution rank into 9 bins: [1, 10], [11, 20], [21, 40], [41, 80], [81, 160], [161, 320], [321, 640], [641, 1280], [1281, 1500] and no rank (no rank means that the institution is not recognized in U.S. News Ranking Database). $R_{ri} = 1$ if the highest institution rank affiliated with a paper $i$ is in bin $r$ and $R_{ri} = 0$ otherwise.
- $F_{fi}$: $F_{fi}$ indicates fixed effects that account for the field-cohort where the author $i$ belongs to. For example, 'Economics' and '2001' is one field-cohort pair indicating all authors in Economics whose first publication years are 2001. Therefore, $F_{fi} = 1$ if an author $i$ belongs to the field-cohort $f$ and $F_{fi} = 0$ otherwise.
- $L_{li}$: $L_{li}$ indicates fixed effects that account for the location (country) of an author $i$. Therefore, $L_{li} = 1$ if an author $i$ belongs to the location $l$ and $L_{li} = 0$ otherwise.

**Table S12** Relationship between matched scientists' citation count and follower count. In model (1), both scientists' citation and follower counts are normalized by their field average. In model (2), both scientists' citation and follower counts are normalized by their field-cohort average.

| Variable | Model (1) DV = norm. follower by field | Model (2) DV = norm. follower by field-year |
|---|---|---|
| Norm. citation by field | 0.073*** (0.011) | |
| Norm. citation by field-year | | 0.057*** (0.012) |
| Controls† | Y | Y |
| Observations | 386,542 | 386,542 |
| R Squared | 0.11 | 0.07 |

†The detail of control variables and fixed effects can be found in section **4.3**.
*, $p < 0.05$; **, $p < 0.01$; ***, $p < 0.001$.

**Table S12** shows that scientists' follower counts are significantly, positively correlated with their citation counts. This result, which also appears in raw data (**Figure 4a**), is shown here to persist when controlling for an author's number of publications, institution rank, field-cohort, and country. The results remain robust when we normalize scientists' citation and number of followers by field-year average (model (2)).

### 6.3 Scientists' Tweeted Paper Impact

**Figure 4b** considers the relationship between follower counts the citation impact of the papers they tweet. We further consider such relationships using fixed-effect ordinary least squares regression as below.

$$y_i = \beta_c c_i + \beta_s s_i + \sum_p \beta_p P_{pi} + \sum_r \beta_r R_{ri} + \sum_f \beta_f F_{fi} + \sum_l \beta_l L_{li} + \epsilon_i \quad (12)$$

**Dependent Variable**: The dependent variable $y_i$ measures the average tweeted paper impact (See Section **S1.2.5** and Equation (3)) by an author $i$.

**Predictors of Interest:** The independent variable $c_i$ indicates an author $i$ normalized citation by field. The independent variable $s_i$ indicates an author $i$ number of followers. We also consider normalizations by field and year as alternatives.

**Control Variables:** We also include several other explanatory variables to control for other possible predictors. The definitions of those control variables can be found in Section **S4.3**.

**Table S13** investigates whether scientists with high follower counts and citation counts are more likely to tweet papers with high impact. Model (1) confirms that both number of followers and citation of scientists are significantly and positively predictive of tweeted papers' impact. In addition, when we normalize number of followers and citation by field-cohort average, we have very similar results.

**Table S13** Relationship between matched scientists' citation/number of followers and tweeted paper impact.

| Variable | Model (1) DV = tweeted paper impact | Model (2) DV = tweeted paper impact |
|---|---|---|
| Norm. follower by field | 0.40*** (0.020) | |
| Norm. citation by field | 0.34*** (0.021) | |
| Norm. follower by field-year | | 0.40*** (0.018) |
| Norm. citation by field-year | | 0.34*** (0.019) |
| Controls† | Y | Y |
| Observations | 358,569 | 358,569 |
| R Squared | 0.15 | 0.16 |

†The detail of control variables and fixed effects can be found in section **4.3**.
*, p < 0.05; **, p < 0.01; ***, p < 0.001.

## 6.4 Percent of Tweeted Papers from the Same Field or Institution

We examine the relationship between matched scientists' citation and the percent of tweeted papers from the same field using a fixed-effect ordinary least squares regression as below.

$$y_i = \beta_c c_i + \beta_s s_i + \sum_p \beta_p P_{pi} + \sum_r \beta_r R_{ri} + \sum_a \beta_a A_{ai} + \sum_f \beta_f F_{fi} + \sum_l \beta_l L_{li} + \epsilon_i \quad (13)$$

**Dependent Variable:** The dependent variable $y_i$ measures the percent of tweeted papers from the same field by an author $i$.

**Predictors of Interest:** The independent variable $c_i$ indicates an author's normalized citation by field. The independent variable $s_i$ indicates an author's number of followers.

**Control Variables:** We also include several other explanatory variables to control for other possible predictors. The definitions can be found in Section **S4.3**.

In this section, we further examine the connection between scientists' follower and citation counts with their likelihood of tweeting papers from their own fields and institutions. We exclude self-promoted papers when calculating the proportion of same-field papers. **Table S14** shows that scientists' follower counts are positively correlated with the likelihood of tweeting papers from their own field. By contrast, scientists with high impact appear, if anything, less likely to mention papers from their own fields. When we normalize a scientist's citation and number of followers by field-cohort average, the results are very similar (see model (2)). We find similar patterns when considering the percent of tweeted papers from the same institution, as presented in **Table S15**.

**Table S14** Relationship between matched scientists' citation/number of followers and percent of tweeted same-field papers.

|  | Model (1) | Model (2) |
|---|---|---|
| Variable | DV = % of same-field papers | DV = % of same-field papers |
| Norm. follower by field | 0.098*** (0.0081) |  |
| Norm. citation by field | -0.021 (0.014) |  |
| Norm. follower by field-year |  | 0.096*** (0.0087) |
| Norm. citation by field-year |  | -0.024* (0.011) |
| Controls† | Y | Y |
| Observations | 331,404 | 331,404 |
| R Squared | 0.18 | 0.18 |

†The detail of control variables and fixed effects can be found in section **4.3**.
*, $p < 0.05$; **, $p < 0.01$; ***, $p < 0.001$.

**Table S15** Relationship between matched scientists' citation/number of followers and percent of tweeted same-institution papers.

|  | Model (1) | Model (2) |
|---|---|---|
| Variable | DV = % of same-institution papers | DV = % of same-institution papers |
| Norm. follower by field | 0.017*** (0.0023) |  |
| Norm. citation by field | -0.0073 (0.0055) |  |
| Norm. follower by field-year |  | 0.016*** (0.0022) |
| Norm. citation by field-year |  | -0.0080 (0.0046) |
| Controls† | Y | Y |
| Observations | 331,404 | 331,404 |
| R Squared | 0.10 | 0.10 |

†The detail of control variables and fixed effects can be found in section **4.3**.
*, $p < 0.05$; **, $p < 0.01$; ***, $p < 0.001$.

# 7 Analysis of Facebook

In this section, we replicate our main analysis using the Facebook mentions of scientific articles. the results are broadly similar to our main findings using X/Twitter data.

## 7.1 Impact and Novelty of Papers Mentioned on Facebook

**Figure S28** repeats the analysis of **Figure 2** but now using instead of Twitter. **Figure S28a** examines paper impact for papers mentioned in mainstream media and on Facebook. We see that Facebook coverage favors high-impact research relative to the baseline, emphasizing works above the 60th percentile of the citation distribution. These results are very similar to those found using Twitter mentions. Further, they continue to show that social media emphasize high-impact works compared to the baseline, but less so than mainstream media.

**Figure S28b** considers the novelty percentile of papers covered by mainstream media and Facebook. Both mainstream media and Facebook emphasize novel papers compared to baseline rates in science, with papers above the 75th percentile of novelty being overrepresented. The pattern for Facebook is very consistent with X/Twitter covered papers shown in **Figure 2b**. Despite significant differences in the number of papers covered on X/Twitter and Facebook, our analysis of the papers covered on Facebook aligns with our findings from X/Twitter.

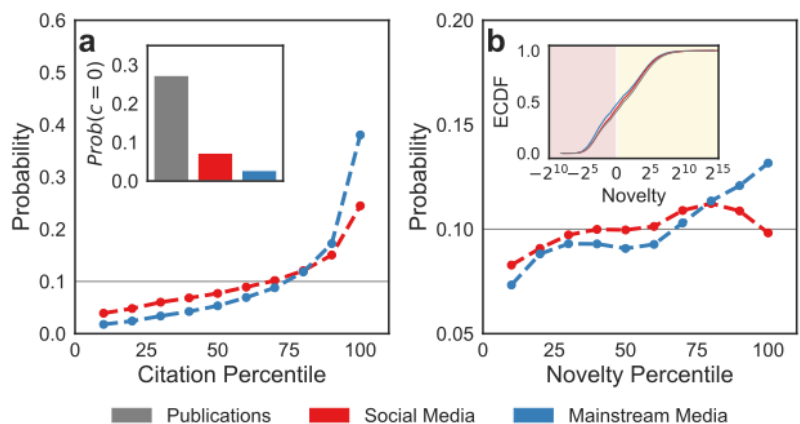

**Figure S28 Paper Impact/Novelty Reported by the Media and Facebook.** Gray/red/blue lines represent the probability density distributions of paper impact or paper novelty for all journal articles/the Facebook reported articles/the mainstream media reported articles. (a) Paper impact is measured by the normalized final citation by field-year, which is converted to percentiles. The x-axis indicates the citation percentile for papers with at least 1 citation ($c \geq 1$). The y-axis quantifies the density. The inset figure reports the percent of papers with zero citation among all journal articles (gray bar), the social media reported articles (red bar), and the mainstream media reported articles (blue bar). (b) Paper novelty measures the degree to which a paper combines past knowledge in a new way (*6*), which is also converted to percentiles. The x-axis indicates the novelty percentile. The y-axis quantifies the probability density. The inset figure represents the novelty z score's (*6*) cumulative density distribution. The z score shows how common a journal pairing is as compared to chance. Negative z scores indicate novel combinations of past knowledge, while positive z scores indicate conventional knowledge combinations.

## 7.2 Heterogeneity of Facebook Covered Papers

**Figure 3** in the manuscript shows that X/Twitter provides more even coverage of scientific papers across various fields, institutions, journals, and teams with different team compositions. Here, we demonstrate that Facebook covered papers show consistent patterns with X/Twitter mentioned papers.

**Figure S29a** examines the likelihood of a paper being reported by mainstream media or mentioned on Facebook within each field of study. Our findings indicate that Facebook provides a more balanced distribution of coverage across scientific disciplines compared to mainstream media. This pattern is consistent with our observations on X/Twitter, although papers are substantially more likely to be covered on X/Twitter than on Facebook on average.

**Figure S29b** examines rate at which papers are mentioned by mainstream media or on Facebook across different institutional ranks. We can see that mainstream media consistently demonstrates a higher report rate for papers from top universities, while the report rate by Facebook is flatter across different institutional ranks.

**Figure S29c** examines the cumulative distributions of the rank of institutions affiliated with papers reported by mainstream media and Facebook. Research from elite universities is disproportionately reported in mainstream media compared to both their share of publications (light grey line) and citations (dark grey line). In contrast, Facebook coverage is less concentrated on elite institutions than mainstream media, while it remains more skewed than citations. The results are broadly consistent with observation of X/Twitter in **Figure 3c**.

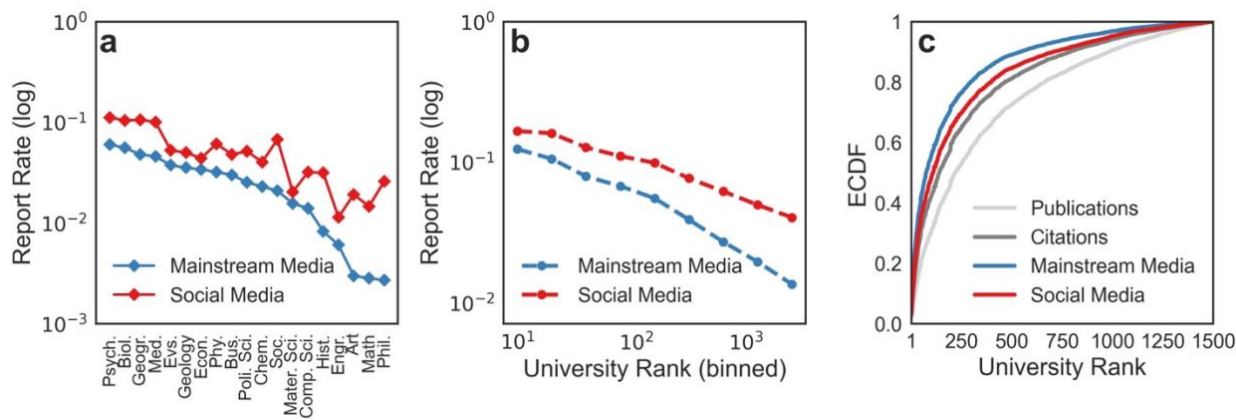

**Figure S29 Coverage of Fields and Institutions**. (a) Plots the probability of reporting on a given field (y-axis) for 19 major topic areas by Facebook (red) and mainstream media (blue). (b) Plots the percent of papers reported on by institutional rank and media source. (c) Plots the empirical cumulative distribution institutional rank for media reported papers (highest rank is 1) where light-grey/grey/red/blue lines represent the ECDF for total publications (light gray), total citations (gray), Facebook mentions (red) and mainstream media mentions (blue).

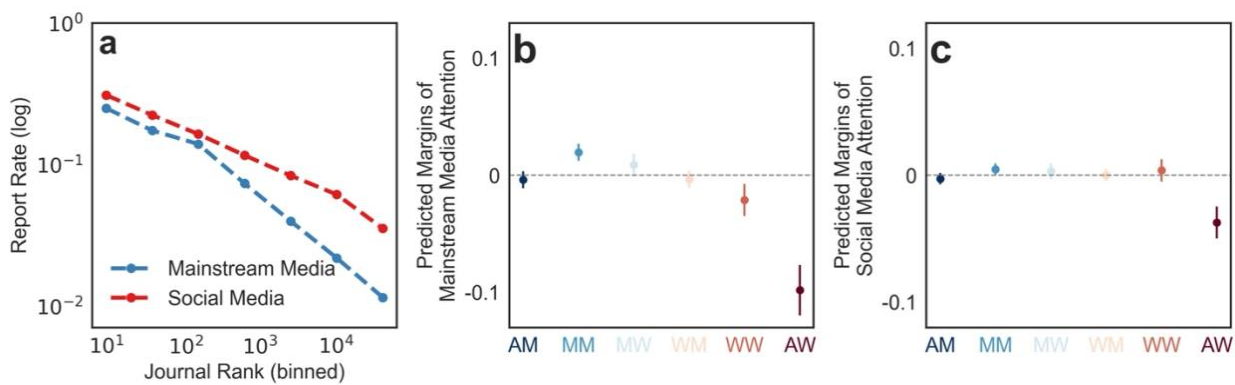

**Figure S30 Journals and Team Composition**. (a) The percent of articles reported by media by journal rank (exponentially binned). (b and c) Present predicted margins of media attention (b) and Facebook attention (c) based on the team's demographic composition measured in six categorical buckets: all women teams (aw), teams with women first/last authors (ww), teams with woman first author and man last author (wm), teams with man first author and woman last author (mw), teams with men first/last authors (mm), and all men teams (am) respectively.

**Figure S30a** examines scientific journals. We see that mainstream media and Facebook coverage strongly emphasize highly-ranked journals, but Facebook provides substantially more coverage of lower-ranked journals. This comparison is broadly similar to the results examining X/Twitter mentions (**Figure 3d**), noting, as before, that X/Twitter coverage has a much higher reporting rate in general.

**Figure S30b** and **S30c** further consider analysis of team composition. We repeat the analysis in **Figure 3e** and **3f** but now using Facebook mentions instead of X/Twitter mentions. The results for Facebook mentions, and in particular the lower report rate for all women teams, are similar to the findings for X/Twitter mentions.

## 7.3 Regression Analysis of Field and Facebook Attention

Similar to our analysis in section **S4.1.1**, we run fixed-effect ordinary least squares regressions predicting Facebook mentions, following the same specification of regression (7). The dependent variable is the number of Facebook mentions.

The predicted margins of field fixed effects are presented in **Figure S31**. The 19 fields are organized in the same order as **Figure 3a** to enable a direct comparison, and the findings are consistent. The variations in margins across different fields on Facebook are notably less pronounced than those observed in mainstream media, yet similar to X/Twitter.

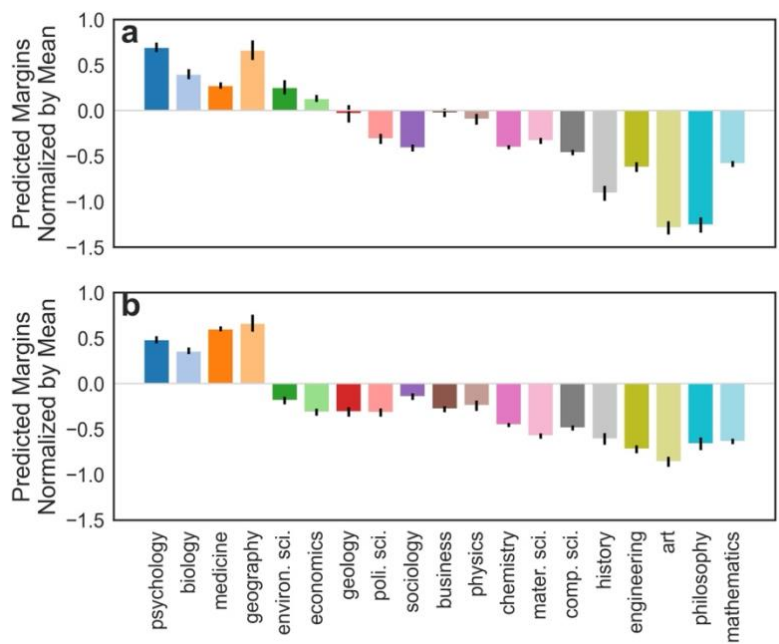

**Figure S31 Fields as Predictors of Mainstream Media and Facebook Mentions**. (a) Margins of field in predicting mainstream media mentions. The x-axis indicates each field. The y-axis quantifies margins of fixed effect for 19 scientific fields. (b) Margins of field in predicting Facebook mentions.

## 7.4 Regression Analysis of Institutional Rank and Facebook Attention

**Figure S32b** and **S32c** consider fixed-effect ordinary least squares regressions, following the same specification of regression (8), to predict paper mentions on Facebook. The gap in normalized margins in mainstream media coverage is approximately twice that observed on Facebook. This finding is consistent with our observations about X/Twitter in **Figure S13**.

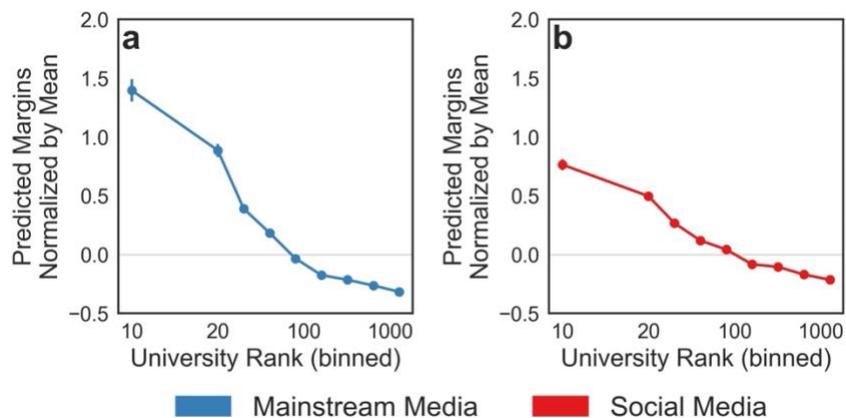

**Figure S32 Institutional Rank in Predicting Mainstream Media and Facebook Mentions**. (a) Margins of institutional rank in predicting mainstream media mentions. The x-axis indicates 9 exponential bins of institutional rank (see Section **S2.1**). Smaller values indicate higher ranks. The y-axis quantifies the margins of each institutional rank level. (b) Margins of institutional rank in predicting the chance of Facebook mentions.

## 7.5 Regression Analysis of Journal Rank and Facebook Mentions

We further conduct the fixed-effect ordinary least squares regression, as in **S4.3.1**, to examine Facebook attention across journals.

Both mainstream media coverage (**Figure S33a**) and Facebook coverage (**Figure S33b**) have a strong propensity towards reporting papers from journals with high Scimago journal ranks. However, the orientation on the highest ranked journals is approximately twice as large in mainstream media as observed on Facebook. These findings are similar to those for X/Twitter.

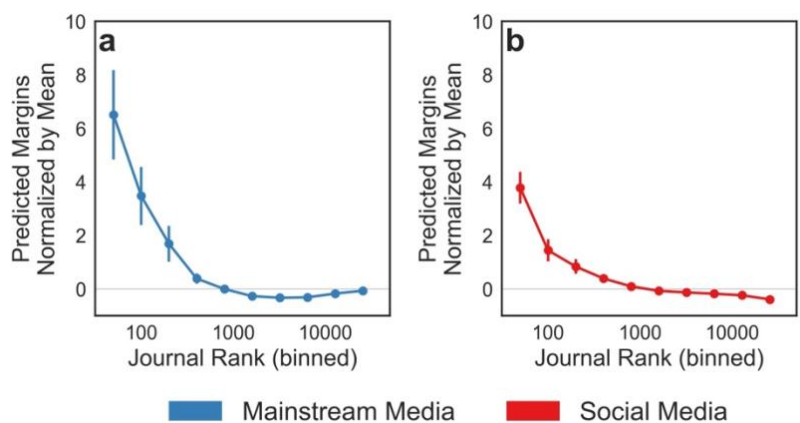

**Figure S33 Journal Rank in Mainstream Media and Facebook Mentions**. (a) Margins of journal rank in predicting mainstream media mentions. The x-axis indicates 10 exponential bins of journal rank. Smaller values indicate higher ranks. The y-axis quantifies the margins of each journal rank level. (b) Margins of journal rank in predicting Facebook mentions.

# 8 Textual Analysis of Abstracts, Media, and Social Media Content

In this section, we analyze the text of paper abstracts, news articles, and tweets that report on scientific research.

## 8.1 Semantic Embedding of Paper Abstracts

We trained a word2vec model on the abstracts of 33 million papers published between 2000 and 2022 in the OpenAlex database, generating 200-dimensional word embeddings. Our main sample comprises 20.9 million journal articles published between 2012 and 2019, of which approximately 17 million include abstracts available in OpenAlex. These abstracts were embedded in the same 200-dimensional vector space using the trained word2vec model.

To visualize the semantic distribution of these papers, we applied incremental principal component analysis (PCA) to reduce the 200-dimensional embeddings to 50 dimensions, followed by t-SNE to project the vectors into a two-dimensional semantic space. **Figure S34** presents the resulting map, showing all papers in our sample (light gray), those covered by mainstream media (blue), and those covered by social media (pink). Unlike the subsampled visualization in **Figure 1** of the main text, **Figure S34** displays the full sample; for clarity, **Figure 1** includes only a 10% random subsample of all papers and those covered by mainstream and social media. The results are consistent across both visualizations: papers covered by social media vastly outnumber those reported by mainstream media. Moreover, social media coverage spans the semantic space more broadly and evenly, suggesting a wider topical reach and a more democratized dissemination of scientific knowledge compared with the concentrated focus of mainstream media coverage.

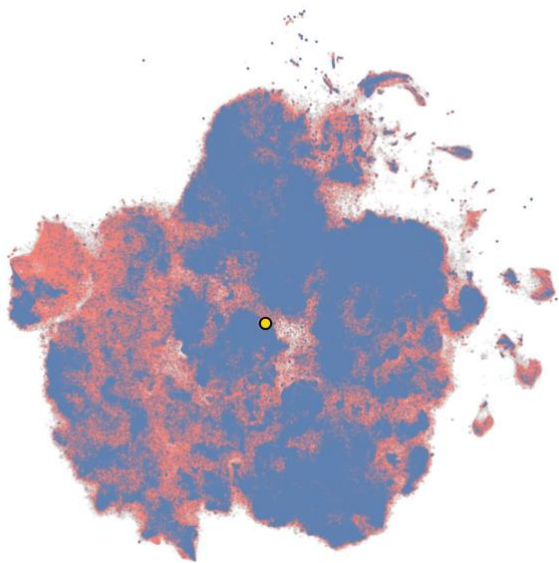

**Figure 34 A Two-dimensional Semantic Map of Scientific Papers (Full Sample).** A two-dimensional semantic map of scientific papers in our sample (light gray), those covered by mainstream media (blue), and those covered by social media (pink; X/Twitter) illustrates the scale difference between the two forms of coverage. Of the 20.9 million articles analyzed, approximately 17 million include abstracts available in OpenAlex. We embedded these papers in a 200-dimensional vector space using a word2vec model trained on roughly 30 million abstracts published between 2000 and 2022, and projected them into two dimensions using incremental PCA for dimensionality reduction followed by t-SNE for visualization.

## 8.2 Measuring Sentiment in Tweets and News Articles

To examine engagement between media discourse and scientific research, we conducted sentiment analysis of news articles and tweets referencing papers in our main sample. For Twitter data, we included original tweets, quote tweets, and replies, excluding retweets because their content largely duplicates the originals. This yielded 18.6 million tweets (34% of all tweets referencing our sample) for text analysis. For news coverage, text was available for 2.06 million of 3.16 million articles (65.1%) scraped from online sources as of 2021; the remaining 34.9% were inaccessible from the provided links.

### 8.2.1 Sentiment Analysis of Tweets
With the rapid growth of social media platforms, sentiment analysis of online content—particularly tweets—has become a key approach for quantifying public reactions. Early systems relied on distant supervision and lexicons, using emoticon- or hashtag-labeled corpora to train feature-rich SVMs and incorporating resources such as the NRC Hashtag Sentiment Lexicon (*8, 9*). Deep learning methods further advanced the field: Tang et al. (*10*) introduced sentiment-specific word embeddings (SSWE) trained on distantly labeled tweets, outperforming feature-engineered baselines, while DeepMoji, pretrained on 1.2 billion emoji-labeled tweets (*11*), achieved strong transfer performance on sentiment and emotion tasks involving short, informal text. The advent of Transformer-based language models in 2018 (*12*) revolutionized sentiment analysis by capturing nuanced linguistic cues. BERT and RoBERTa models fine-tuned on Twitter data set new performance benchmarks, and the TweetEval framework (*13*) unified evaluation across Twitter tasks, demonstrating further gains from continued pretraining on in-domain tweets—underpinning widely used models such as *twitter-roberta-base-sentiment*. Most recently, large language models (LLMs) have enabled zero- and few-shot sentiment classification without task-specific training. Benchmark studies report that GPT models achieve competitive accuracy (*14, 15*) but can exhibit polarity over-assignment, with GPT-4 showing sensitivity to prompt design, underscoring the importance of prompt formulation (*16*).

In this study, we use both GPT-4 and TweetEval—a RoBERTa-based model trained on 124 million tweets and fine-tuned on multiple large sentiment datasets—to analyze the sentiment of tweets referencing scientific papers (*13*). A recent evaluation (*14*) comparing GPT-3.5 and GPT-4 Turbo for sentiment analysis of Twitter posts about heated tobacco products reported accuracies of approximately 57% and 77%, respectively. The TweetEval model has demonstrated a comparable classification accuracy of 72.8% in different datasets. Our main analysis employs the GPT-4o-mini model with a prompt defining positive, negative, and neutral sentiment categories. For robustness, we also apply GPT-4o-mini using a prompt without predefined definitions across all 18.6 million tweets. In addition, we include a state-of-the-art RoBERTa-based model ([cardiffnlp/twitter-roberta-base-sentiment-latest](cardiffnlp/twitter-roberta-base-sentiment-latest), updated in 2022) for comparison.

**Figure S35** shows the distributions of sentiment scores for tweets referencing scientific papers. The top panel (**Figure S35**, **a–c**) displays positive/negative/neutral sentiment scores generated by the GPT-4o-mini model using a prompt that defines each sentiment category. The middle panel (**Figure S35**, **d–f**) presents corresponding distributions from GPT-4o-mini using a prompt without predefined categories. The bottom panel (**Figure S35**, **g–i**) shows results from TweetEval (*13*), a RoBERTa-based model trained on 124 million tweets. Across all three models, tweets classified as neutral have the largest share, followed by positive and negative tweets. The GPT-4o-mini model without predefined sentiment categories, however, diverges notably from the other two models—a difference consistent with prior findings (*16*), which we discuss in a later section.

Using GPT-4o-mini with a prompt defining each sentiment category, 72.3% of tweets are classified as neutral (neutral sentiment $\geq 0.5$). When the sentiment definitions are removed from the prompt, the share of neutral tweets drops sharply to 47.9%—a decrease of nearly 25 percentage points. In contrast, the fine-tuned RoBERTa-based TweetEval (13) model classifies 78.3% of tweets as neutral, showing strong agreement with GPT-4o-mini when sentiment categories are explicitly defined. Positive sentiment represents the second largest share across all three models. Based on GPT-4o-mini with predefined sentiment categories, 16.7% of tweets are classified as positive (positive sentiment $\geq 0.5$), whereas removing the sentiment definitions nearly doubles this share to 36.9%. The TweetEval model identifies 13.0% of tweets as positive, again showing close alignment with GPT-4o-mini when definitions are included. Finally, GPT-4o-mini with defined categories classifies 13.2% of tweets as negative, compared to 14.2% when prompt definitions are omitted and 7.7% using TweetEval. Overall, all three models consistently indicate that tweets covering scientific papers are most likely to convey neutral sentiment. These results are further validated by visualizing 18.6 million tweets in a three-dimensional sentiment space using a ternary density plot (**Figure S37**). Results from the three models described above are shown in **Figure S37 (a–c)**. In all cases, most tweets cluster near the neutral vertex, with a secondary concentration near positive sentiment; tweets with predominantly negative sentiment form the smallest share of the sample. This pattern is consistent across the three models used in our analysis.

Our findings in **Figure S35** are aligned with prior studies based on smaller datasets. Friedrich et al. (*17*) reported that 94.8% of tweets referencing scientific findings were neutral in a sample of 1,000 tweets, and a subsequent larger study (*18*) found 81.7% neutral, 11.0% positive, and 7.3% negative tweets (n = 487,610). These findings are further confirmed by the work of (*19*). Drawing a significantly larger dataset (n = 18.6 million), our results present consistent findings.

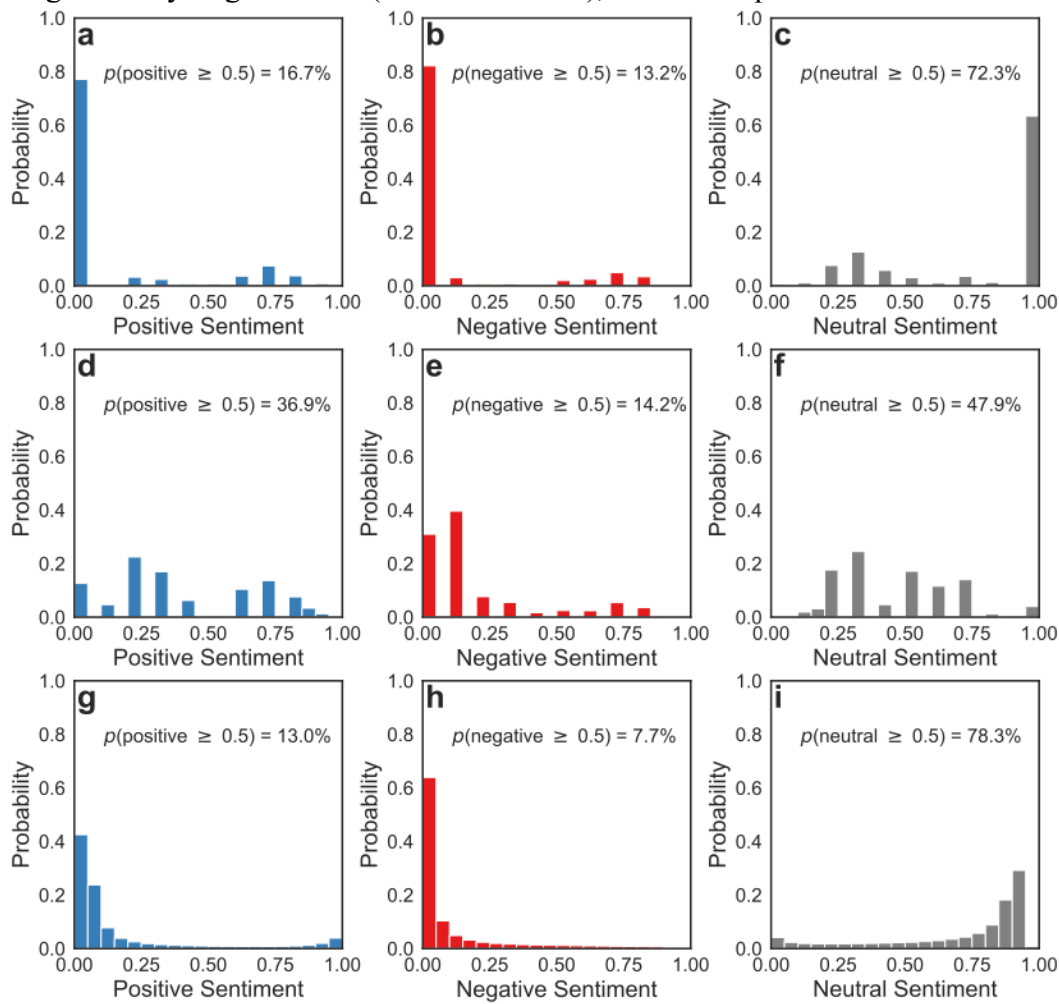

**Figure S35 Sentiment Score Distributions of Tweets.** (a–c) Distributions of sentiment scores assigned by GPT-4o-mini when using a prompt that includes definitions of sentiment categories. (d–f) Distributions of sentiment scores assigned by GPT-4o-mini when using a prompt without predefined sentiment categories. (g–i) Distributions of sentiment scores assigned by the TweetEval model (*13*).

There is a clear discrepancy between the results of GPT-4o-mini when sentiment categories are explicitly defined in the prompt and when they are not. As shown in **Figure S35**, this difference primarily arises from neutral tweets being misclassified as positive and, less frequently, as negative. **Figure S36** illustrates several real examples where GPT-4o-mini without predefined categories disagrees with both GPT-4o-mini using defined sentiment categories and the TweetEval model. The tweets shown in **Figure S36** are neutral in tone, typically disseminating scientific findings or promoting research papers without expressing overt sentiment. However, GPT-4o-mini without predefined categories often misclassifies these neutral tweets as strongly positive or negative. This pattern is consistent with prior literature (*16*). Tweets are short, context-dependent, and often ambiguous. Unlike news articles that provide longer contextual cues, LLMs analyzing tweets may over-rely on surface features—such as positive or negative words, emoticons, or hashtags—and misinterpret nuance. This limitation is especially relevant when applying LLMs in zero-shot settings to infer sentiment in tweets reporting on science. Abraham et al. (*20*) show that even small changes in prompt phrasing can cause substantial swings in labeling accuracy across social science tasks. Similarly, Li et al. (*21*) demonstrate that incorporating explicit knowledge into prompts improves sentiment interpretation, particularly for fine-grained tasks. Several other studies also emphasize that clearly defining category boundaries enhances the robustness and consistency of LLM-based classification (*22, 23*). Despite these discrepancies, our main conclusion—that the majority of tweets express neutral sentiment—remains robust.

We also observe strong agreement between GPT-4o-mini with a prompt defining each sentiment category and the TweetEval model. TweetEval is a RoBERTa-based model trained on 124 million tweets and has demonstrated state-of-the-art performance in tweet-level sentiment classification (https://github.com/cardiffnlp/tweeteval). Bucher *et al.* (*24*) showed that smaller LLMs—such as RoBERTa-based models—fine-tuned on task-specific labeled data can outperform zero-shot generative models like GPT-4 across a range of text classification tasks, including sentiment analysis. Several evaluations of GPT-3.5 and GPT-4 in zero-/few-shot settings (*14, 16*) likewise report competitive but not consistently superior performance, underscoring that domain-specific fine-tuning remains difficult to surpass without careful prompting or supervision (*16*). These findings explain the strong agreement between GPT-4o-mini with defined sentiment categories and the TweetEval model, and further justify our use of GPT-4o-mini with explicit category definitions as the main model in this study.

| Tweet | GPT-4o-mini (with definitions) | GPT-4o-mini (without definitions) | TweetEval |
|---|---|---|---|
| takashi haramura published methods for invasive species control are transferable across invaded areas. as the corresponding author in plos one (public library of science, usa). 3 nov, 2017 https://t.co/uv5oodkkem | Neutral: 1.0 | Positive: 0.7 | Neutral: 0.90 |
| membrane translocation of cytoplasmic ifn-r2 is a critical step for the activation of #macrophage innate response against intracellular bacterial #infection https://t.co/pwrqoqsopw https://t.co/w0bmb3e7jm | Neutral: 1.0 | Positive: 0.7 | Neutral: 0.74 |
| from our highly cited collection: long-read technologies what they are, why they are useful, and how they are being used in medical research, clinical diagnostics, and therapeutics. https://t.co/ftw6srj28n https://t.co/vf9tjtpf7a | Neutral: 0.7 | Positive: 0.7 | Neutral: 0.66 |
| differences in facial expressions during positive anticipation and frustration in dogs awaiting a reward https://t.co/odgkhmoknz | Neutral: 1.0 | Positive: 0.6 | Neutral: 0.77 |
| mimicking biological functionality with polymers for biomedical applications. https://t.co/e7izforqry #tumorimmuno | Neutral: 1.0 | Positive: 0.6 | Neutral: 0.90 |
| the last frontiers of wilderness: tracking loss of intact forest landscapes from 2000 to 2013 https://t.co/q6laicdkwm | Neutral: 1.0 | Positive: 0.7 | Neutral: 0.74 |
| spinal cord gray matter atrophy correlates with multiple sclerosis disability http://t.co/sdjlnxsxnp via @feedly | Neutral: 1.0 | Positive: 0.7 | Neutral: 0.69 |

**Figure S36 Examples of Tweets Misclassified by GPT-4o-mini When the Prompt Omits Predefined Sentiment Categories.** Most discrepancies relative to the other two models stem from neutral tweets being labeled as positive (predominantly) and, less frequently, as negative.

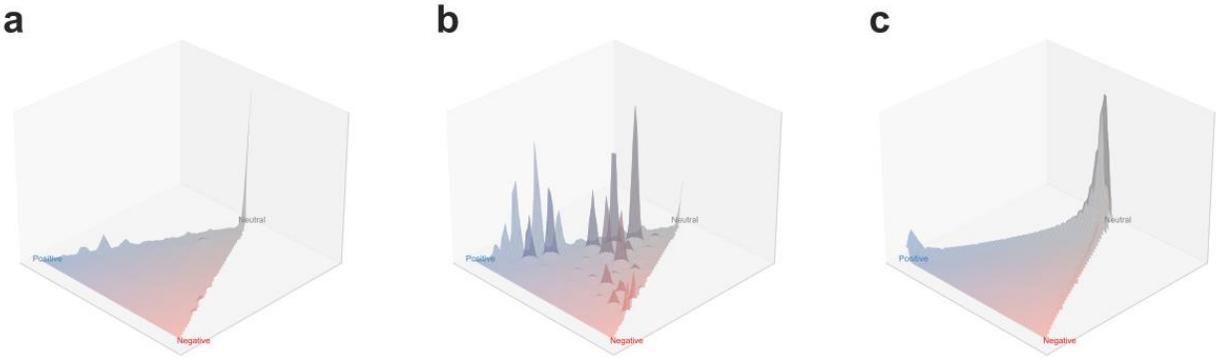

**Figure S37 Distribution of Tweets in Three-Dimensional Sentiment Space.** (a) Three-dimensional ternary density plot of sentiment scores assigned by GPT-4o-mini using a prompt that includes explicit definitions of sentiment categories. (b) Three-dimensional ternary density plot of sentiment scores assigned by GPT-4o-mini using a prompt without predefined sentiment categories. (c) Three-dimensional ternary density plot of sentiment scores assigned by the TweetEval model.

### 8.2.2 Sentiment Analysis of News Articles

For news articles, lexicon- and rule-based systems established early baselines. SentiWordNet 3.0 (*25*) provided word-level polarity priors widely used in sentiment pipelines, while the Loughran–McDonald lexicon (*26*) introduced finance-specific tone categories—such as negative, litigious, uncertainty, and modal—that became standard in analyses of news and disclosures. The rule-based VADER system (*27*) offered a lightweight and generalizable baseline. Malo et al.'s phrase-structure approach (*28*) contributed expert-labeled data and methods tailored to news semantics, forming a lasting benchmark for evaluating news sentiment.

As in tweet-level analysis, machine learning approaches were later applied to classify news sentiment. Hierarchical Attention Networks (HAN) (*29*) introduced an attention hierarchy for multi-paragraph news text, serving as a baseline before transformer architectures became dominant. Fine-tuning generic BERT and RoBERTa models on news corpora improved accuracy, but domain-specific variants such as FinBERT (*30*) achieved further gains through adaptive pretraining on financial text—consistent with the finding (*31*) that continued in-domain pretraining enhances performance in news and finance sentiment classification. Most recently, large language models (LLMs) have achieved competitive zero- and few-shot performance on news sentiment tasks. BloombergGPT (*32*), a domain-specialized model, reported strong results on financial sentiment benchmarks, while empirical studies (*33-35*) show LLMs (i.e., Llama 3, GPT-3.5 and GPT-4) outperforming FinBERT in sentiment analysis.

In our work, we use both GPT-4 and Llama 3 (Llama 3 8B instruct, https://huggingface.co/meta-llama/Meta-Llama-3-8B-Instruct) to analyze the sentiment of news articles reporting scientific papers. Our main analysis employs the GPT-4o-mini model with a prompt defining positive, negative, and neutral sentiment categories (consistent with the prompt used in the sentiment analysis of tweets). For robustness, we also apply GPT-4o-mini using a prompt without predefined definitions across all 2.06 million news articles. In addition, we report the results of Llama 3 for comparison.

**Figure S38** shows the distributions of sentiment scores for news articles. The top panel (**Figure S38**, **a–c**) displays the positive, negative, and neutral sentiment scores generated by the GPT-4o-mini model using a prompt that defines each sentiment category. The middle panel (**Figure S38**, **d–f**) presents corresponding distributions from GPT-4o-mini using a prompt without predefined sentiment categories. The bottom panel (**Figure S38**, **g–i**) shows results from the Llama 3 8B model. All three models indicate that most news articles reporting on science express positive sentiment. This main result aligns with prior literature done on smaller and more selective samples (*36-38*).

As a robustness check, we used different prompts and LLMs and found consistent evidence. Using the fact that the three sentiment probabilities sum to one, we visualize 2.06 million news articles in a three-dimensional sentiment space using a ternary density plot (**Figure S39**). Results from the three models described above are shown in **Figure S39 (a–c)**. In all cases, most news articles cluster near the positive vertex, with a secondary concentration near negative sentiment; articles with predominantly neutral sentiment form the smallest share of the sample. This pattern is consistent across the GPT-4o-mini model with defined sentiment categories, GPT-4o-mini without predefined categories, and the Llama 3 8B model [4].

Our observations align with several existing studies. Prior work (*23, 39*) has noted that large language models (LLMs) tend to overpredict the majority class, and that clearly defining category boundaries improves robustness (*22, 23*). This pattern helps explain why GPT-4o-mini without sentiment definitions and Llama 3 8B classify more news articles as positive compared with GPT-4o-mini using defined sentiment categories. Despite issues of polarity bias, LLMs continue to show competitive performance on sentiment classification tasks (*33-35, 40*). Moreover, Zhang *et al.* (*40*) found that simple sentiment classification tasks are more robust and stable than more complex ones—such as aspect-based sentiment analysis—under different prompt formulations. Hu *et al.* (*22*) demonstrated that "knowledgeable prompt-tuning" improves both effectiveness and stability in zero- and few-shot settings. These findings support our choice of GPT-4o-mini with explicit sentiment category definitions for the main analysis.

Our findings in **Figure S38** and **Figure S39** also align with prior research on media reporting of science. Sumner *et al.* (*41*) showed that exaggeration in academic press releases predicts exaggeration in subsequent news coverage. Previous studies have also documented that science news often skews optimistic. For example, a content analysis of 774 print and online articles on personalized or precision medicine (*36*) found that approximately 82% of stories present these advances positively, emphasizing benefits while downplaying concerns. Likewise, Bubela and Caulfield (*37*), analyzing 627 news articles covering 111 genetic research papers, found that most systematically highlighted benefits and underreported risks—consistent with a general tendency toward positive sentiment in science communication. Garvey and Maskal (*38*) conducted a sentiment analysis of AI news coverage and similarly found that media discourse is not predominantly negative but largely neutral or positive. Our study extends this line of work by providing one of the first large-scale sentiment analyses of news articles reporting on scientific research, reinforcing these patterns using a substantially larger dataset.

---

[4] Using GPT-4o-mini with a prompt defining each sentiment category, 53.2% of articles are classified as positive (positive sentiment ≥ 0.5). When the sentiment definitions are removed from the prompt, the share of positive articles rises to 61.3%—an increase of 8.1%. Applying an older model, Llama 3 8B, further increases the proportion of positive articles to 72.0%, 18.8% higher than GPT-4o-mini with defined categories. In contrast, the proportion of negative news articles shows much smaller variation across models. Using GPT-4o-mini with a prompt defining each sentiment category, 20.3% of articles are classified as negative (negative sentiment ≥ 0.5). Removing the sentiment definitions increases this share slightly to 22.3%—a difference of 2%. The Llama 3 8B model classifies 19.6% of articles as negative. Finally, GPT-4o-mini with a prompt defining each sentiment category classifies more news articles as neutral (14.4% of articles having neutral sentiment ≥ 0.5) compared to 2.6% by the GPT-4o-mini without sentiment category definitions and 1.2% by Llama 3 8B. Despite these differences, all three large language models (LLMs) consistently indicate that most news articles reporting on science convey positive sentiment.

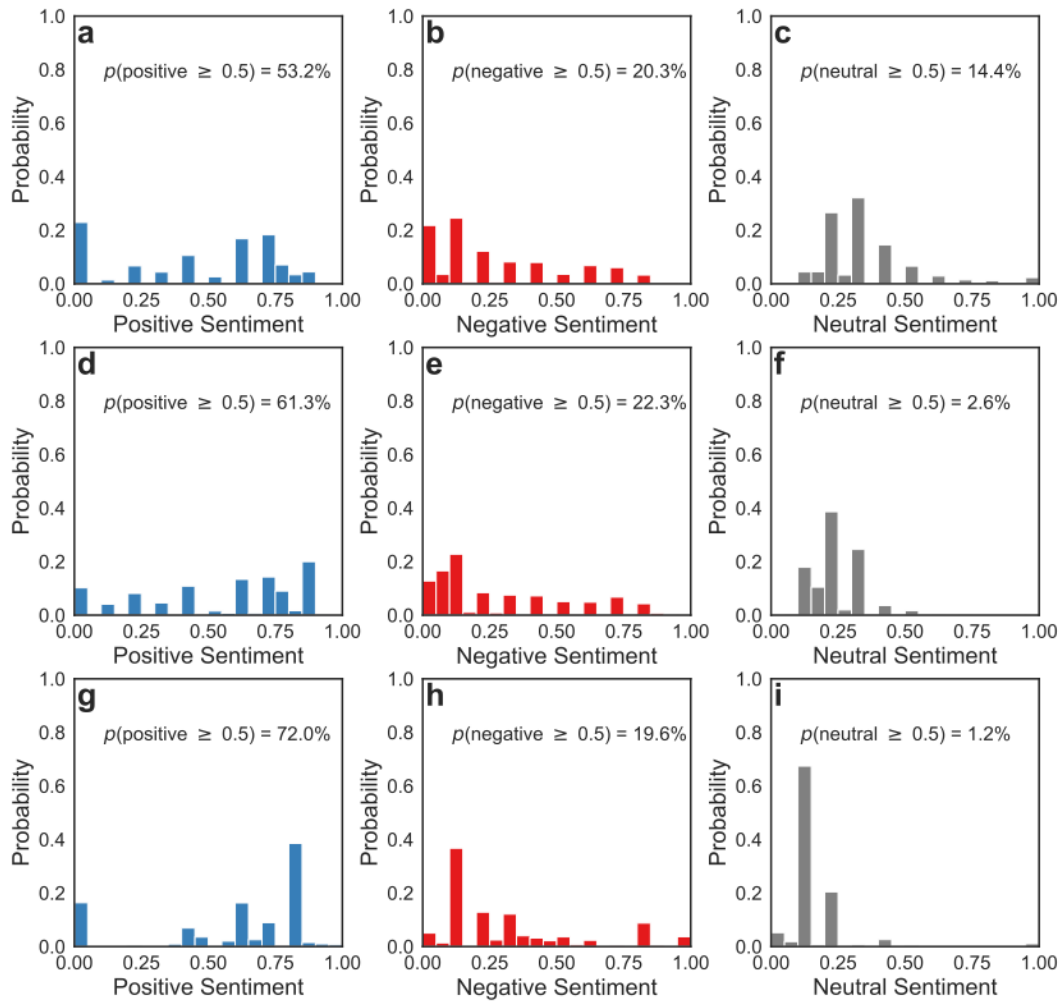

**Figure 38 Sentiment Score Distributions of News Articles.** (a–c) Distributions of sentiment scores assigned by GPT-4o-mini when using a prompt that includes definitions of sentiment categories. (d–f) Distributions of sentiment scores assigned by GPT-4o-mini when using a prompt without predefined sentiment categories. (g–i) Distributions of sentiment scores assigned by the Llama-3-8B model.

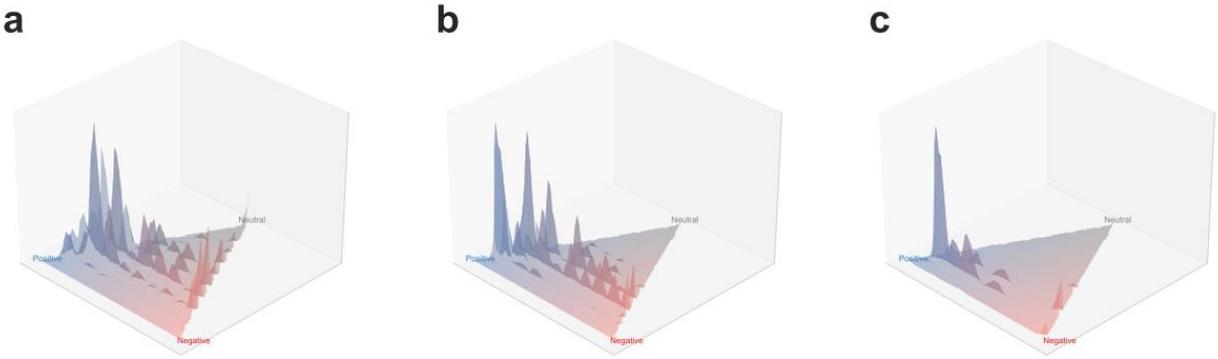

**Figure 39 Distribution of News Articles in Three-Dimensional Sentiment Space.** (a) Three-dimensional ternary density plot of sentiment scores assigned by GPT-4o-mini using a prompt that includes explicit definitions of sentiment categories. (b) Three-dimensional ternary density plot of sentiment scores assigned by GPT-4o-mini using a prompt without predefined sentiment categories. (c) Three-dimensional ternary density plot of sentiment scores assigned by the Llama-3-8B model.

## 8.3 Identifying Frames in Tweets and News Articles

To further examine how tweets and news articles portray science, we identified frames and associated keywords within the three sentiment groups classified by GPT-4o-mini using the prompt with defined sentiment categories, aiming to uncover the dominant narratives and themes through which science is represented.

Pastorino et al. (*42*) evaluated GPT-3.5, GPT-4, and FLAN-T5 on detecting framing bias in news headlines under zero-shot, few-shot, and explainable prompting conditions, finding that GPT-4 achieved strong performance. Similarly, Alonso et al. (*43*) demonstrated that large language models can reliably identify media frames, performing comparably to traditional supervised classifiers when guided by well-designed prompts. Building on these insights, we carefully calibrated our prompt in GPT-4o-mini to include detailed definitions of candidate frames. We then conducted frame and keyword detection on a randomly selected 5% subsample of the three sentiment groups (positive, negative, and neutral) for both news articles and tweets.

We make several observations. First, the number of frames associated with each news article is, on average, significantly greater than that of an individual tweet. A news article contains an average of 4.5 frames, with positive, negative, and neutral articles averaging 4.5, 4.7, and 4.5 frames, respectively. By contrast, tweets exhibit far fewer frames, averaging only 1.4 overall. Positive, negative, and neutral tweets contain, on average, 1.6, 1.9, and 1.3 frames, respectively. Both positive and negative tweets tend to include more frames than neutral tweets: the median number of frames is one for neutral tweets and two for both positive and negative tweets.

Second, the frames used in news articles and tweets differ substantially. Across all three sentiment groups (positive/negative/neutral), *Caution/Uncertainty/Limitations* is consistently the most common frame in news articles, aligning with prior research (*44, 45*) showing that linguistic uncertainty/hedging are established norms in science reporting and shape audience perceptions of credibility. In our 5% subsample, 99.3% of positive, 99.9% of negative, and 99.2% of neutral news articles include this frame. The associated keywords overlap considerably across sentiment groups and include terms such as "may," "preliminary," "limitations," "caveats," "uncertainties," and "no causal claims." This pervasive framing reflects a common journalistic practice of emphasizing caution and methodological restraint when reporting scientific findings (*44, 45*).

Beyond this near-universal framing, news articles of different sentiment categories exhibit distinct thematic emphases. Positive news articles frequently adopt *Social Progress/Benefits* (96.7%) and *Breakthrough/Hype/Novelty* (86.9%) frames. The *Social Progress/Benefits* frame highlights the contributions of scientific discoveries to societal advancement, often using keywords such as "quality of life," "health benefits," and "could help patients." The *Breakthrough/Hype/Novelty* frame, by contrast, underscores novelty and impact through terms like "game-changing," "groundbreaking," "major advance," and "breakthrough." This pattern is consistent with prior studies showing that health and science communication often employ benefit framing and, at times, hype or exaggeration driven by source press releases (*41, 46*). Negative news articles, in contrast, commonly use *Risk/Safety/Pandora's Box* (67.7%) and *Social Progress/Benefits* (66.1%) frames. The *Risk/Safety/Pandora's Box* frame invokes concerns about unintended consequences, using keywords such as "adverse effects," "ethical concerns," "potential risks," "raises risks," and "unintended consequences." In negative articles, the *Social Progress/Benefits* frame takes on a more cautionary tone, emphasizing "sustainability," "safety," and "awareness." This aligns with observations of the existence of risk framing in science and technology coverage (*47*). Finally, neutral news articles most often adopt *Neutral Reporting/Factual Summary* (76.7%) and *Social Progress/Benefits* (76.9%) frames—an expected pattern consistent with their informational intent.

Third, the framing patterns observed in tweets differ markedly from those in news articles. Unlike news coverage—where *Caution/Uncertainty/Limitations* appears almost universally—no single frame dominates across tweet sentiment groups. Positive tweets most frequently adopt *Breakthrough/Hype/Novelty* (47.6%), *Neutral Reporting/Factual Summary* (47.2%), and *Social Progress/Benefits* (37.4%) frames. The *Breakthrough/Hype/Novelty* frame highlights important scientific advances using keywords such as "breakthrough," "novel," "amazing," "discovery," and "exciting." Although these differ slightly from those found in news articles, they convey similar messages of enthusiasm and scientific significance. This pattern is consistent with prior work (*48*) showing that novel information tends to diffuse more rapidly and widely on social media, helping explain the frequent use of the *Breakthrough/Hype/Novelty* frame in positive tweets. The *Neutral Reporting/Factual Summary* frame reflects straightforward dissemination of new research, with frequent use of terms like "published," "article," "research," "new paper," "evidence," and "findings."

In contrast, negative tweets are most commonly linked with the *Caution/Uncertainty/Limitations* frame (85.7%). However, unlike in news articles, this frame in tweets relies on a distinctly different vocabulary—featuring terms such as "caution," "risks," "increased risk," "concerns," and "challenges." Whereas journalists typically use this frame to emphasize methodological restraint, social media users invoke it to highlight perceived dangers or shortcomings of scientific work. Negative tweets also use *Conflict/Controversy/Strategy* (23.1%) and *Risk/Safety/Pandora's Box* (21.9%) frames, reflecting more contentious or cautionary engagement with scientific content. Finally, most neutral tweets adopt *Neutral Reporting/Factual Summary* (84.5%) frame as expected, with a small portion of neutral tweets also using Caution/Uncertainty/Limitations (23.5%) and Breakthrough/Hype/Novelty (8.6%) frames.